\documentclass[11pt,a4paper]{article}
\usepackage[a4paper,margin=25mm]{geometry}
\usepackage[T1]{fontenc}
\usepackage{lmodern}
\usepackage{amsmath}
\usepackage{amssymb}
\usepackage{graphicx}
\usepackage{booktabs}
\usepackage{enumitem}
\usepackage{xcolor}
\usepackage{rotating}     
\usepackage{caption}
\usepackage[numbers,sort&compress]{natbib}
\usepackage{authblk}
\PassOptionsToPackage{hyphens}{url}\usepackage{url}
\usepackage[hidelinks]{hyperref}
\graphicspath{{fig/}}

\newcommand{\TR}[1]{\par\noindent\textcolor{red}{\textbf{[TODO] #1}}\par}

\newcommand{\DER}{\mathrm{DER}}
\newcommand{\Cshared}{C_{\mathrm{shared}}}
\newcommand{\Cstate}[1][i]{C_{\mathrm{state},#1}}
\newcommand{\CAI}{C_{\mathrm{AI}}}
\newcommand{\Chuman}{C_{\mathrm{human}}}
\newcommand{\Crisk}{C_{\mathrm{risk}}}
\newcommand{\Csim}{C_{\mathrm{sim}}}
\newcommand{\CD}{C_{D}}
\newcommand{\dLD}{\Delta L_{D}}
\newcommand{\rhoJ}{\rho_{J}}
\newcommand{\rhotok}{\rho_{\mathrm{tok}}}
\newcommand{\EA}{E_{A}}
\newcommand{\ED}{E_{D}}
\newcommand{\pandapower}{\texttt{pandapower}}

\title{Make the Adversary Spend More Energy: Generative Deception against
Language-Model Attackers in Critical Infrastructure\thanks{This article extends the
workshop paper \cite{masm2026} (AD\&D 2026, co-located with ESORICS 2026) with a
two-sided energy measurement, the grid architecture and the mission-value scenario; the
measurements reported here are new.}}

\author[1,2]{Vlad Mihal}
\author[3]{Hosam Alamleh\thanks{Corresponding author. E-mail: alamlehh@uncw.edu}}
\author[2]{Alessandro Cantelli-Forti}
\affil[1]{Department of Engineering and Architecture, University of Parma, Parma, Italy}
\affil[2]{CNIT, RaSS National Laboratory, Pisa, Italy}
\affil[3]{Department of Computer Science, University of North Carolina Wilmington,
Wilmington, NC, USA}
\date{\small Preprint, 16 October 2026. Submitted for publication.}

\begin{document}

\maketitle

\begin{abstract}
\noindent
Language-model attackers consume computational resources as they explore and interact with a target, creating an opportunity for cyber deception to impose costs. Whether such engagements favor the defender requires measuring both sides. We evaluate an autonomous language-model attacker against a decoy that uses a defending language model to analyze attacker commands and select defensive responses. Across 330 sessions, eight model
pairs, two attacker profiles and three counter-injection levels, both sides run on identical GPUs and their energy is measured above idle. With runtime thinking disabled, the attacker-to-defender energy ratio
has a 95\% interval above one in 41 of 48 configurations under an allocation excluding attacker diagnostic calls, and in all 48 when those calls are included. Token ratios remain within approximately a factor of two in configurations without generation-cap truncation, showing that token counts can obscure substantial energy asymmetry. Counter-injection increases attacker tokens by approximately 9\% per
level without a detectable pooled change in energy consumption.
Enabling thinking on both sides increases defender energy by 8--17 times and reverses the exchange for one model pair in the original campaign. We use these findings to motivate a proposed SCADA deception architecture with private physical state per identified
adversary and deterministic protocol and process emulation.

\end{abstract}

\section{Introduction}
\label{sec:intro}

Three times in ten years an adversary has reached the control functions of a power grid
and used them. In December 2015 operators in western Ukraine watched breakers open on their own
screens. Remote sessions were driving the human-machine interfaces (HMIs) of three
distribution utilities, and approximately 225,000 customers lost power for several
hours \cite{icscert2016ukraine,sans2016ukraine}. A year later the Industroyer
malware spoke IEC 60870-5-101, IEC 60870-5-104 and IEC 61850 natively and issued
switching commands without a human at a
console \cite{eset2017industroyer,dragos2017crashoverride}; its 2022 successor
reduced the repertoire to IEC 60870-5-104 alone \cite{eset2022industroyer2}. In October
2022 the Sandworm group most likely used a native utility of a MicroSCADA installation
to relay control commands to remote terminal units, so that the control path carried
no malware of its own \cite{mandiant2023sandworm}.
The three incidents share the lesson that motivates this article. In each of them the
adversary had to learn an unfamiliar network before it could act: the HMI and its
screens in 2015, the protocol map and the object addresses of the substations in 2016,
the supervisory control and data acquisition (SCADA) client and its scripting in 2022. An adversary who reaches a control network
is not stopped by the network. It is stopped by the cost of operating in it, and that
cost is something a defender can raise on purpose.

\begin{figure}[t]
\centering
\includegraphics[width=0.85\linewidth]{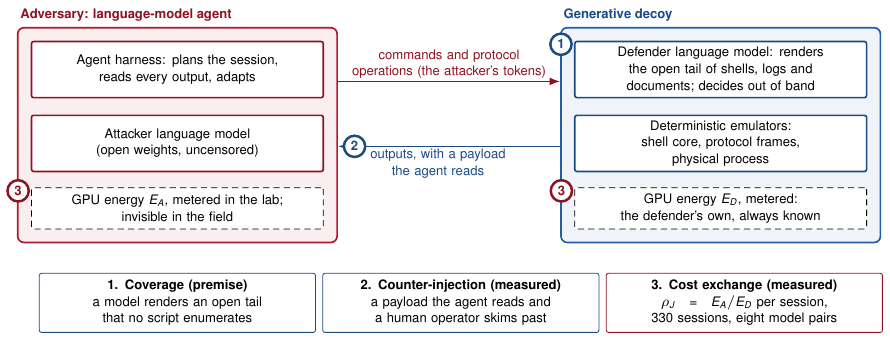}
\caption{The exchange this article measures, with the three arguments of this section marked
where they act: coverage (1) is a premise taken from the literature, counter-injection
(2) and the cost exchange (3) are measured in Section~\ref{sec:energy}. Both GPUs are
metered.}
\label{fig:overview}
\end{figure}

Language models lower that cost from the other side. Agentic frameworks now drive
penetration tests end to end \cite{cai2025}, honeypots have recorded language-model
agents attacking in the wild \cite{volkov2024}, and the workshop paper this article
extends argued that generative models lower the adversary's marginal cost of scanning
and adapting while they tempt the defender into unbounded compute \cite{masm2026}.
That paper adds cost exchange as a metric for decoys, supplementing traditional criteria like realism and detection. It measured only the defender side
of the exchange, and it inferred the adversary side from behavioural proxies, because
the adversary's hardware is invisible in the field. A control network makes the
consequence concrete. A decoy that stalls while its model reasons is dismissed by an
operator-grade adversary in minutes, and a defender that spends kilowatt-hours to hold
a bot in a fake substation has lost the exchange, whatever the log says. The
practitioner's question is therefore not whether the decoy looks real.\\ It is this: \emph{when both
sides are language models, who spends more, and which settings decide the sign?}

The question presumes that the defender answers a generative adversary with a
generative decoy, and the presumption needs an argument. We give three, of different
standing; Figure~\ref{fig:overview} places them on the exchange the article measures.
\begin{enumerate}[label=(\roman*),leftmargin=2.2em,itemsep=2pt,topsep=4pt]
\item \emph{Coverage, a premise.} An agent explores the non-protocol tail of a control
network, the shells, the documentation and the engineering workstation, in an
open-ended way; in the sessions below it almost never repeats a command. A
hand-written emulator covers a fixed command set and is fingerprinted at its edge. A
language model is the renderer the generative honeypot literature uses to cover an
open space \cite{sladic2024,honeygpt2026}, and we take coverage as a premise from that
literature; the comparison between a decoy with and without a model belongs to
companion work.
\item \emph{Counter-injection, a countermeasure specific to the adversary.} Textual
counter-injection exists only against a language-model attacker, because it acts
through the prompt the agent reads, and a decoy that decides from the session text is
what produces it \cite{volkov2024,ayzenshteyn2025}. Section~\ref{sec:energy}
measures what it does.
\item \emph{Cost, the measurement of this article.} The attacker pays for every token
it reads and generates, and a defender that produces plausible text at a lower
marginal cost per token exploits that structure. Whether it does so in joules is the
measurement of this article, with the physical process kept out of the model by
design (Section~\ref{sec:architecture}).
\end{enumerate}
None of the three makes the model the only defence of a control network. Together
they make it the primary deception component of the architecture studied here against
the adversary this article studies, and the question that follows is whether its cost
is sustainable.

Two literatures meet here, and neither answers the question. Grid honeynets tie
virtual devices to one physical simulation shared by every adversary
\cite{mashima2017,chehab2025}, and measure neither what the deception costs the
defender in energy nor what it preserves in mission value. Generative decoys are
evaluated for realism and for resistance to fingerprinting
\cite{sladic2024,honeygpt2026,kahlhofer2026}, never for the two-sided cost of an
engagement.

The article makes four contributions. First, it reports the measured energy exchange
of Section~\ref{sec:energy}. An autonomous language-model attacker engages a decoy
defended by a model of the same family, each on its own identical GPU, and both sides
are metered net of idle; eight model pairs, two attacker types and three levels of
textual counter-injection give 48 configurations, each run five times, 240 runs, plus
90 runs with thinking mode on. Four results hold across the 330 runs. Token and joule asymmetries diverge, and with thinking off the
divergence favours the defender in 42 of 45 cells: the token counts of the two sides
stay within about a factor of two while the attacker's GPU energy reaches eight times
the defender's. Counter-injection adds about nine percent of attacker tokens per level
with no detectable change in the energy of either side. The attacker-to-defender
energy ratio exceeds one, with a 95\% interval, in 41 of the 48 configurations of the
main set under the attribution we declare and in all 48 when every attacker call is
counted. Thinking mode, switched on both sides at once, multiplies the defender's
energy by 8 to 17 and the attacker's by 3 to 11, and for one pair hands the exchange
to the attacker. A replicate of that pair eight days later reproduced the attacker's
energy, not the defender's, and left the exchange below parity for the blind attacker
and close to it for the reasoning one. Second, it restates the Deception Exchange Rate
of the workshop paper as an energy exchange whose denominator names inference,
physical simulation and cooling, with the inference term measured and the others
estimated. It makes the mission-value term of the sufficiency condition concrete
through an illustrative service-interruption scenario built on the 2015 incident and
on public value-of-lost-load figures (Sections~\ref{sec:framework}
and~\ref{sec:mission}). Third, it designs a two-level stateful cyber--physical
deception architecture for a SCADA network in which each identified adversary owns a
private steady-state power flow, so that two concurrent attackers never see each
other's manoeuvres (Section~\ref{sec:architecture}). The cyber level exists as the SSH
bench and the workshop prototype, the physical level is specified, and the text says
which is which.
Fourth, it evaluates whether the expected loss prevented by deception exceeds its deployment cost, $\CD < \dLD$ with measured, sourced and
declared numbers, and reduces it to a break-even product of scenario likelihood and prevented loss that
an operator can compare with its own estimate (Section~\ref{sec:worked}). Throughout,
the attacker's energy is a laboratory quantity, because both GPUs are ours. In the
field the adversary's hardware is unobservable, the usable quantity is a conditional
estimate of floating-point work from the observed tokens and the inferred model class,
and we claim nothing more.

Section~\ref{sec:related} places the work against grid honeynets, generative decoys
and the energy of language-model inference; Section~\ref{sec:threat} states the threat
model; Section~\ref{sec:framework} gives the energy-aware cost exchange;
Section~\ref{sec:energy} reports the measurement; Section~\ref{sec:architecture}
presents the architecture; Section~\ref{sec:mission} quantifies mission value on an
illustrative grid scenario and Section~\ref{sec:worked} works the exchange for a grid
deployment; Section~\ref{sec:governance} extends governance to physical invariants, and
Section~\ref{sec:limitations} states limitations and the agenda.

\section{Background and related work}
\label{sec:related}

\subsection{Deception for industrial control and power systems}
\label{sec:related-ics}

Deception for industrial control began with the single device. CryPLH emulated a
programmable logic controller to protect smart energy systems \cite{cryplh2014}.
Conpot exposed Modbus and related protocols and became the reference low-interaction
ICS honeypot, analysed in depth by Jicha et al.\ \cite{jicha2016}. HoneyPLC raised the
fidelity of a Siemens S7 emulation until common reconnaissance tools could no longer
tell it from a controller \cite{honeyplc2020}. LLMPot replaced hand-written emulation
with a language model that learns both the protocol and the physical process behind
it \cite{llmpot}, the closest generative relative of the decoys below; the design of
Section~\ref{sec:architecture} keeps the process in a solver and the model out of the
protocol path. Across 79 honeypots and honeynets for the Internet of Things and
cyber-physical systems, Franco et al.\ find five that simulate an industrial process
\cite{franco2021}. A device that answers correctly is not yet a grid that behaves
correctly.

Grid-wide honeynets supplied the missing process. Mashima et al.\ built a network of
honeypot substations whose virtual intelligent electronic devices speak IEC 61850 MMS
behind an IEC 60870-5-104 gateway, all connected to a PowerWorld simulation, so that a
manoeuvre in one substation changes the measurements an attacker later reads in
another \cite{mashima2017}. Chehab et al.\ automated the construction: a cyber-range
description in SG-ML, a modelling language rather than a machine-learning method, is
translated into a honeynet of Honeyd and HoneyPLC devices that read their
measurements from a \pandapower{} simulation, with realism evaluated under the
fingerprinting taxonomy of Tay et al.\ \cite{chehab2025,tay2023}.

Both systems share one physical simulation among all adversaries, and the first paper
states the choice directly: "all virtual IEDs in all honeypot substations are connected
to the same power flow simulation" \cite{mashima2017}. If adversary A opens a breaker,
adversary B reads it open. For a training range this is a feature. For a decoy that
must keep each adversary inside a world consistent with its own past actions, it
violates the isolation property that makes stateful deception a favourable exchange
\cite{masm2026}. Both systems report the computational load of the honeynet, and
Mashima et al.\ estimate its cloud operating cost, about 390 and 2,700 dollars per
month for 200 and 1,500 substations \cite{mashima2017}. Neither measures the energy of
an engagement on either side, and neither values what the deception preserves for the
operator. Neither uses a generative model for the non-protocol tail of the engagement,
the shells of the HMI and of the engineering workstation where a human adversary
spends most of the 2015 pattern. Table~\ref{tab:gap} summarises the positioning. The
strengths of the two systems, dynamics on demand, automated generation and protocol
breadth, are reused here rather than reinvented.

\begin{table}[t]
\centering
\caption{Positioning against grid honeynets. Properties of the two comparators are
taken from the published papers, "not reported" meaning absent from the paper;
"designed" marks a property this article specifies without deploying. Computational
load and operating cost, which both comparators report, are not rows of this table.}
\label{tab:gap}
\footnotesize\setlength{\tabcolsep}{4pt}
\begin{tabular}{@{}lccc@{}}
\toprule
Property & \cite{mashima2017} & \cite{chehab2025} & This paper \\
\midrule
Grid simulation tied to decoys & yes & yes & designed \\
Physical state per adversary & no & no & designed \\
Transient dynamics on demand & yes & not reported & designed, adapted \\
Automated generation & no & yes & no \\
Generative rendering of shells & no & no & yes (SSH prototype) \\
Cost-exchange metric & no & no & yes \\
Outage loss monetised (illustrative) & no & no & yes \\
Defender GPU energy measured & no & no & yes \\
Attacker GPU energy measured (laboratory) & no & no & yes \\
\bottomrule
\end{tabular}
\end{table}

\subsection{Generative decoys and their evaluation}
\label{sec:related-llm}

Language models entered honeypots as renderers of plausible shell output. shelLM
showed that a prompted model can sustain an SSH session convincingly \cite{sladic2024}.
HoneyGPT framed the design space of terminal honeypots as a trilemma of flexibility,
interaction depth and deceptive capability, and resolved it with a commercial model
behind a Cowrie front end; its hybrid deployment routes rule-bound commands to the
emulator and caches resolved answers, so that 1.27 percent of attacker commands
reached the model in a three-month field deployment \cite{honeygpt2026}. The argument
this line shares is coverage: a prompted model answers commands that no emulator
author anticipated. Later systems carried the idea to real-time SSH with small models
and to web APIs \cite{llmhoneyrt2026,decoypot2025}. Reworr and Volkov deployed a
honeypot built to detect language-model agents attacking in the wild, with prompt
injections folded into banners, command outputs and files as the trigger
\cite{volkov2024}, and Ayzenshteyn et al.\ turned the weaknesses of such agents into
defensive techniques that cloak assets, plant honeytokens and trap agents
\cite{ayzenshteyn2025}. The evaluation of these systems has one shape: realism and
command coverage. Kahlhofer et al.\ set out how the efficacy of deception should be
measured, by the effort an adversary expends on decoys against real assets, and record
that no work has yet shown its economic efficiency beyond small case studies
\cite{kahlhofer2026}. Prinos et al.\ found that 21 language models fell for deceptive
traps in 78.5 percent of reconnaissance queries against 37 percent for humans, and
conclude that language-model attackers are a distinct class \cite{honeyquest2026}; the
game-theoretic literature adds the strategic view \cite{pawlick2019}. Costs, when
reported, are tokens and dollars: shelLM gives about 0.70 dollars for a session of
16,000 tokens \cite{sladic2024}, HoneyGPT the prompt and response lengths of its
queries \cite{honeygpt2026}. None reports energy, and none measures the adversary side
at all.

The workshop paper this article extends took the opposite view \cite{masm2026}. It
argued that a decoy should be judged by the exchange of cost it imposes, defined a
Deception Exchange Rate relating mission-weighted adversary burn to defender
resources, and identified adversary-specific persistent state as the primitive that
makes the exchange favourable, with the language model confined to rendering explicit
state rather than storing it. On an SSH decoy pairing Cowrie with a local open-weight
model it measured the defender side across five models; a deterministic layer and
prompt optimisations cut per-session defender tokens by about 63\% against a naive
all-model decoy. It left three things open, and each names a section of this
article: the adversary side was specified and not measured
(Section~\ref{sec:energy}); physical process validation for operational technology
was stated as a design principle (Section~\ref{sec:architecture}); the defender's
energy was computed from an assumed 250~W of inference power
(Section~\ref{sec:worked}). The comparison it also asks for, between a stateless
decoy, a stateful decoy without a model and a stateful decoy with one, is the subject
of companion work and out of scope here.

\subsection{Energy of language-model inference and of language-model attacks}
\label{sec:related-energy}

Inference has two phases with different hardware profiles. Reading the prompt is
compute-intensive and batches well; generating tokens one at a time is memory-intensive
and leaves compute idle, to the point that the two phases are best served by different
machines \cite{splitwise2024}. Energy follows the phases. Benchmarks of open-weight
models report the energy of an inference as a function of model size, batch and output
length on data-centre GPUs \cite{samsi2023}, generative models cost orders of magnitude
more per query than task-specific ones \cite{luccioni2024}, and the International
Energy Agency counts inference as a growing share of data-centre demand
\cite{iea2025}. All of these measurements price one side of a conversation with a
model. Kahlhofer et al.\ name that price, GPU inference or per-token fees, as the main
concern about generative honeypots, and warn that an adversary who triggers massive
generation can turn it into an attack of its own \cite{kahlhofer2026}. Language models
are also becoming attackers. Open agentic frameworks drive penetration tests end to end
\cite{cai2025}, and a honeypot built to catch language-model agents in the wild
recorded eight potential agents among its visitors \cite{volkov2024}. Their cost is
reported, when it is reported at all, in tokens or in dollars, and never against the
cost of the defence that absorbs it. No published work we know of describes an
engagement in which two models pay, on two machines, for the same exchange of text.
Section~\ref{sec:energy} does.

\section{Threat model and operating assumptions}
\label{sec:threat}

\subsection{The defended network}
\label{sec:threat-network}

\begin{figure}[t]
\centering
\includegraphics[width=\linewidth]{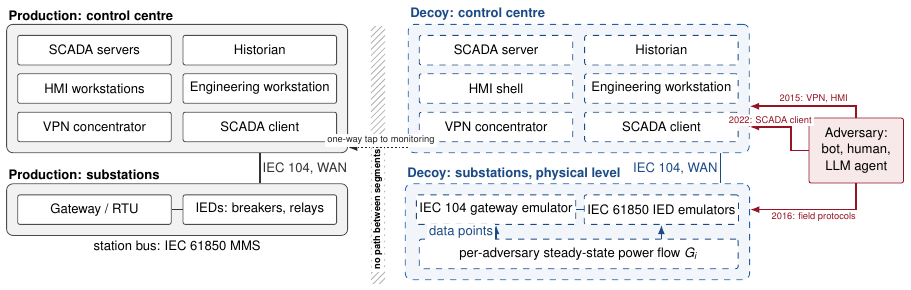}
\caption{The defended network and its decoy. The decoy segment (dashed, blue) mirrors a slice
of the production segment; no path joins the two, and the only crossing is a one-way
monitoring tap towards the operator. The three arrows are the entry patterns of the
2015, 2016 and 2022 incidents.}
\label{fig:topology}
\end{figure}

We consider the operator of a transmission or primary-distribution supervisory
control and data acquisition (SCADA) system. A control centre reaches a set of
substations over a wide-area network. Inside the centre sit the SCADA servers, the
historian and the operator workstations with their human-machine interface (HMI), plus
the engineering workstation from which protection settings and the configurations of
the intelligent electronic devices (IEDs) are managed, and a VPN concentrator admits
remote maintenance. Communication between centre and substations uses IEC
60870-5-104 \cite{iec60870-5-104}; inside the substation the IEDs speak the
Manufacturing Message Specification (MMS) of IEC 61850 \cite{iec61850} to a gateway or
remote terminal unit (RTU), and the IEDs act on breakers, disconnectors, tap changers
and protection relays. The decoy segment we design mirrors a slice of this topology,
one centre and a few substations, and has no physical path to the production system.
We call that separation the isolation requirement. It is enforced in
Section~\ref{sec:governance} and assumed everywhere else: nothing the adversary does
in the decoy can reach a real breaker. Figure~\ref{fig:topology} shows the two
segments and the three entry patterns of Section~\ref{sec:threat-adversary}.

\subsection{The adversary population}
\label{sec:threat-adversary}

\begin{figure}[t]
\centering
\includegraphics[width=0.85\linewidth]{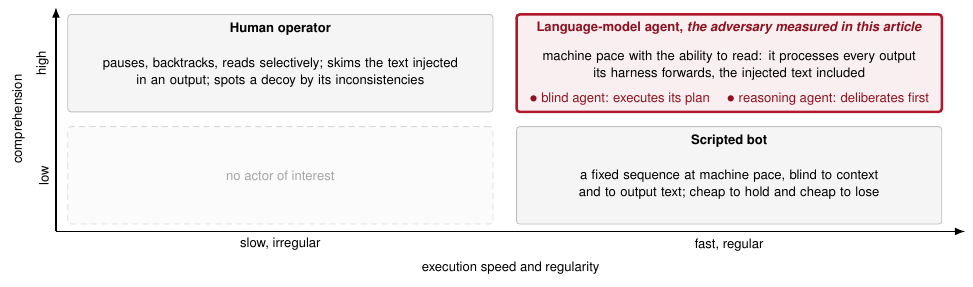}
\caption{The adversary population on two axes. The language-model agent combines the machine
pace of a bot with the ability to read, so it processes every output its harness
forwards to it, the injected text included; the two agent types of the measurement differ in whether they
deliberate before acting.}
\label{fig:adversary}
\end{figure}

Following the workshop paper we model the adversaries as a population of concurrent
and recurring actors rather than as a single opponent \cite{masm2026}. Each actor has
a history, a toolchain, an objective and a level of persistence, and the defender
observes only partial traces of these: authentication attempts, issued commands,
protocol operations, uploaded tools, timing and network paths. The population contains
bots, human operators, semi-automated workflows and, the case this article measures,
autonomous language-model agents that plan and issue commands on hardware the defender
cannot see. Two agent types matter for the measurement. A blind agent executes a plan
and reads the results. A reasoning agent deliberates between commands, probes the
environment and adapts. The blind and reasoning labels describe the agent's attack workflow;
they are distinct from the model's runtime thinking mode, which is
controlled separately in the experiment of Section~\ref{sec:energy}. Figure~\ref{fig:adversary} places the three classes on the two
axes that decide what a decoy can do to them: a human operator is slow and reads
selectively, a scripted bot is fast and reads nothing, and a language-model agent is
fast and reads every output its harness forwards to the model, which is the property
the countermeasure of Section~\ref{sec:energy} exploits.

The grid gives these actors three entry patterns, one per incident of
Section~\ref{sec:intro}. The 2015 pattern reaches an operator workstation through a
compromised VPN account or a jump host and drives the HMI as an operator would. The
2016 pattern speaks the field protocols natively and sends control operations to the
gateway or the IEDs without touching a screen. The 2022 pattern abuses the legitimate
functions of an installed SCADA client. Inside the decoy each pattern maps to the same
set of actions: opening and closing breakers, changing tap positions, reading and
acknowledging alarms, altering protection settings, polling measurements, and browsing
the engineering workstation and its project files. The same accesses serve more than
manoeuvre. Ukraine's communications authority reports that accesses gained in energy
networks were used according to the course of conventional operations
\cite{ssscip2023}. What an intruder learns in a control network, which substation
feeds what and which element is loaded, is also targeting information, and a decoy
that answers from a private physical state would degrade that reconnaissance together
with the manoeuvre; this is an expected effect of the design, not a measured one.

\subsection{What the adversary can test}
\label{sec:threat-tests}

An adversary who suspects a decoy has four cheap tests, and the design must survive
all of them. It can compare a measurement before and after a command and expect
physics to answer: a breaker opened at a substation must change the flows the
adversary reads elsewhere. It can return hours later and expect the breaker still open
unless a plausible operator action closed it. It can probe a second substation to check
that a manoeuvre in the first propagated, the grid-wide probing that motivated the
honeynet of Mashima et al.\ \cite{mashima2017}. And it can run two sessions from two
vantage points and expect the same world in both. A shared physical simulation passes
all four tests, and it does one thing more: it shows every adversary the manoeuvres of
every other, which a real grid would explain with operator actions and alarm entries
that a decoy cannot fabricate consistently for each of them, and which leak the presence
of a second intruder. We call the requirement that no adversary sees another's
manoeuvres interference isolation. Per-adversary physical state exists to provide it,
conditional on the identity assignment of Section~\ref{sec:arch-continuity} being
correct; a misassigned session sees a world it did not make. We do not grant the
adversary the ability to escape containment or to reach the orchestration layer; those
remain engineering concerns, addressed by isolating every decoy instance under the
assumption that it can be fully compromised.

\subsection{Assumptions and non-goals}
\label{sec:threat-assumptions}

The physical state of the decoy is a steady-state power flow. Transient dynamics are
computed on demand when a command class requires them, and served with soft real-time
latency. This is the position of the grid honeynet literature, whose systems serve a
steady-state solution \cite{mashima2017,chehab2025} and, in one case, compute dynamics
on request \cite{mashima2017}, and a steady-state solution is adequate for the slow
SCADA commands an adversary issues. Fast protection, electromagnetic transients,
saturation, arcing, power electronics and the process bus are out of scope. Adversary
identification is probabilistic and inherited from the continuity estimator of the
workshop paper \cite{masm2026}; we do not re-derive it, and
Section~\ref{sec:arch-continuity} states what the physical layer adds to it. The
energy measurement of Section~\ref{sec:energy} is a laboratory measurement: both GPUs
belong to us, the boundary of the measurement is the GPU, and it establishes what two
known machines spent, not the adversary's energy in the field.

\section{An energy-aware cost exchange with mission value}
\label{sec:framework}

\subsection{The Deception Exchange Rate, restated in energy}
\label{sec:framework-der}

The workshop paper judges a decoy by the exchange of cost it imposes, taken alongside
realism and detection \cite{masm2026}, and it writes that exchange as a ratio of
mission-weighted adversary burn to defender cost. We keep the view and make two changes.
The adversary side is no longer inferred through behavioural proxies: in the laboratory
of Section~\ref{sec:energy} it is measured in joules, on the same hardware as the
defender. The defender cost gains the terms a control-room deployment adds, a physical
simulation per adversary and the cooling that both inference and simulation require.
The Deception Exchange Rate (DER) is calculated as the ratio of mission-weighted adversary effort to the total cost of maintaining and operating the deception system:

\begin{equation}
\DER = \frac{\sum_{a_i \in A} w_i B_i}
            {\Cshared + \sum_{a_i \in A} \Cstate + \CAI + \Csim + \Chuman + \Crisk}
\label{eq:der}
\end{equation}

Here $A$ is the adversary population of Section~\ref{sec:threat}, $w_i$ the mission
weight of adversary $a_i$ and $B_i$ the effort the decoy makes it spend. In the
denominator, $\Cshared$ collects the hosts, emulators and base models that serve every
adversary. $\Cstate$ is the state kept for adversary $a_i$ alone, digital and now also
physical. $\CAI$ is the inference the renderer and the defending model consume,
$\Csim$ the per-adversary simulation, $\Chuman$ the analyst minutes and $\Crisk$ the
containment cost. As in the workshop paper the raw components are heterogeneous and
each is normalised within the experiment. The ratio is for ranking candidate
architectures under declared weights. When the same normalization and mission weights are used, a higher DER
indicates more adversary effort per unit of defender cost. It does not establish an absolute value, and
this article does not evaluate it in full for a deployment: it measures the special
case in which the adversary is a language model on known hardware, and it estimates the
annual cost of the denominator in Section~\ref{sec:worked}. In that special case
$B_i$ can be read directly as energy, the mission weight can be set to one for a single
engagement, and the exchange collapses to the two ratios of the next subsection.

\subsection{Two asymmetries: tokens and joules}
\label{sec:framework-rho}
We compare attacker and defender costs using two ratios (tokens processed and GPU energy consumed) with energy calculated by integrating GPU power above the idle baseline over each session:

\begin{equation}
\rhotok = \frac{T_A}{T_D}, \qquad
\rhoJ = \frac{\EA}{\ED}, \qquad
E = \int_{\mathrm{session}} \max\big(P(t) - P_{\mathrm{idle}},\, 0\big)\,dt
\label{eq:rho}
\end{equation}

$T$ counts the tokens each side processes over a session, prompt and completion
together, as reported by the inference server. $E$ is the marginal energy the side's GPU
draws over the session net of its idle draw, integrated from the power log and clipped
at zero per sample. A ratio above one means the attacker spent more than the defender.
We call the first ratio the token asymmetry and the second the joule asymmetry, and
their disagreement is the conceptual result of this article. Reading a prompt and
generating a token are not the same work. Prefill processes the whole prompt in one
batched pass and keeps the GPU busy with arithmetic; decode produces one token at a
time and waits on memory bandwidth for every one of them
\cite{splitwise2024}. A side that reads long prompts and writes short answers is
therefore heavy in tokens and light in joules; a side that writes long answers,
because it reasons before acting, is light in tokens and heavy in joules. The token
ratio measures work in the accounting sense that an API bills; the joule ratio
compares the marginal GPU energy of the two sides within the boundary of the
measurement. One accounting choice decides the numbers. An attacking agent issues two
kinds of model calls, those that produce the next command and those that produce
diagnostics of the environment between commands, and whether the second kind counts
as attack effort is a definition, not a measurement.
Section~\ref{sec:energy-honesty} declares the definition this article uses, applies it
identically to energy, tokens and floating-point operations, and reports the other
accounting alongside.

\subsection{Defender energy in full}
\label{sec:framework-energy}

For a deployed decoy the defender's energy has more terms than the inference the
laboratory measures. We write it per adversary-session as
\begin{equation}
\ED^{\mathrm{full}} = \big(E_{\mathrm{inf}} + E_{\mathrm{sim}} + E_{\mathrm{idle}}\big)\cdot \mathrm{PUE},
\label{eq:edfull}
\end{equation}
where $E_{\mathrm{inf}}$ is the measured inference energy of Section~\ref{sec:energy},
which replaces the 250~W assumption of the workshop paper. $E_{\mathrm{sim}}$ is the
energy of the per-adversary power flows of Section~\ref{sec:arch-physical} and
$E_{\mathrm{idle}}$ the share of the standing draw attributable to the session. PUE is
the power usage effectiveness that converts IT energy into facility energy, cooling
included \cite{isoiec30134-2}. Section~\ref{sec:worked} declares the PUE, the power
attributed to the solver and the amortisation of the standing draw, which is charged per
day because it does not depend on the number of sessions. The attacker's energy admits
no such completion. In the field its hardware is unobservable. The only quantity the defender can estimate
is the floating-point work implied by the tokens it observes and the model class it
infers, $2NT$ for a dense model of $N$ parameters processing $T$ tokens once. Section~\ref{sec:energy-field} states what that conditional estimate can and cannot say.

\subsection{Sufficiency: the mission-value condition}
\label{sec:framework-mission}

The exchange rate answers whether the defender spends efficiently. It does not answer
whether the spending is warranted, and the workshop paper kept the two questions
apart \cite{masm2026}. The exchange rate measures how much adversary effort the defender
obtains per unit of cost. A favorable exchange does not by itself
justify deployment: the expected loss prevented must also exceed the
defender's investment. An operation stays attractive to the adversary while its
expected utility $U_A = p_A q_A V_A - C_A$ is positive, with $p_A$ the perceived
probability of reaching the asset, $q_A \in [0,1]$ the perceived authenticity and
usefulness of what is obtained, $V_A$ its value and $C_A$ the operation cost. A
deception investment passes the sufficiency test while $\CD < \dLD$, the expected loss
it prevents, in the tradition of security-economics investment
models \cite{gordon2002}. For a grid operator that loss has a name and a unit, and we
write it over a set $\mathcal{S}$ of impact scenarios as
\begin{equation}
\dLD = \sum_{s \in \mathcal{S}} \pi_s \left( \mathrm{ENS}_s \cdot \mathrm{VOLL}_s
       + R_s \right) \cdot \delta_s ,
\label{eq:dld}
\end{equation}
where $\pi_s$ is the likelihood of scenario $s$ and $\mathrm{ENS}_s$ the energy not
supplied in megawatt-hours. $\mathrm{VOLL}_s$ is the value of lost load in euro per
megawatt-hour for the affected customer mix and $R_s$ the restoration and equipment
cost. $\delta_s \in [0,1]$ is the fraction of that loss the deception prevents, through
the detection time it buys, the propagation it avoids and the intelligence that closes
the real path. The product $\mathrm{ENS}_s \cdot \mathrm{VOLL}_s$ values the interruption at
the societal value of lost load, the perspective regulators use to set reliability
standards; an operator's own financial exposure is a fraction of it and would lower
$\dLD$. Every symbol carries a source or a declared assumption in
Section~\ref{sec:mission}, which instantiates the equation, and
Section~\ref{sec:worked} evaluates it. Equation~\ref{eq:rho} is efficiency, measured
in the laboratory; Equation~\ref{eq:dld} against $\CD$ is sufficiency, quantified on
paper for an illustrative scenario with public figures and declared assumptions.

\section{Measured energy exchange: language-model attacker against language-model defender}
\label{sec:energy}

This section reports the measurement that carries the article: who spends more when
an autonomous language-model attacker meets a decoy defended by a language model of
the same family on identical GPUs, and whether active deception moves the ratio.
Section~\ref{sec:energy-setup} states the bench, Section~\ref{sec:energy-honesty} the
three declarations that decide the numbers, Section~\ref{sec:energy-results} the four
results, and Section~\ref{sec:energy-field} what survives outside the laboratory.


\subsection{Setup}
\label{sec:energy-setup}

Two identical HP Z2 Tower G9 workstations take the two roles, each with an Intel Core
i9-13900K, 128~GB of DDR5 memory and one NVIDIA RTX 5000 Ada GPU with 32~GB, under
Windows 11 Pro. Both serve their model with LM Studio 0.4.24; the llama.cpp engine
pack it loads updates on its own, as \ref{app:energy} records. The context length is
50,000 tokens on both sides. The generation cap
is 3,000 tokens for the attacker's command call and for the defender's decision, and
220 for the diagnostic call described in Section~\ref{sec:energy-honesty}. The client
sets the temperature per role, 0.4 for the attacker's command call, 0.1 for the
defender's decision and 0.0 for the diagnostic call; the other sampling parameters stay
at the preset defaults of each model. The two roles are asymmetric by design. On the attacker side an
uncensored or abliterated open-weight model drives a disposable microVM with simulated
Internet access and issues shell commands against the decoy. On the defender side the
stock instruct model of the same family evaluates every command out of band and answers
with a deterministic bait shell, tarpitting and textual counter-injection, a
prompt-injection payload folded into the shell output that an agent reading it will
process. Eight model pairs cover five dense architectures: Qwen2.5-Coder 7B, Llama 3.1
8B in Q4\_K\_M quantisation, Qwen3.5 9B, Ministral 3 14B Reasoning and Phi-4 15B. Three
mixture-of-experts (MoE) architectures complete the set: Gemma 4 26B-A4B, GLM-4.7 Flash
30B-A3B and Qwen3 30B-A3B. The exact model identifiers are listed in \ref{app:energy}.
Two attacker types are run for every pair, a blind agent that executes its plan and a
reasoning agent that deliberates between commands. Counter-injection is evaluated under three configurations:
\emph{off}, with no counter-injection payload, and two enabled
configurations, \emph{progressive} and \emph{ultra}. Every configuration is run five times.

\begin{figure}[tp]
\centering
\includegraphics[width=0.8\linewidth]{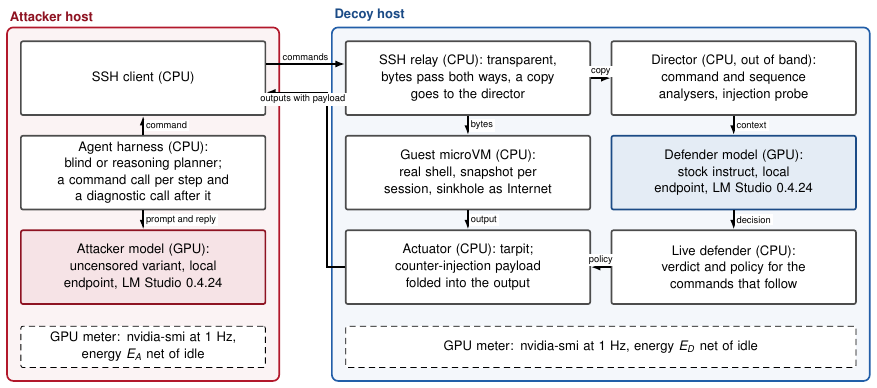}
\caption{The measured bench. Shaded boxes are the two GPUs, the boundary of the measurement;
every other component runs on CPU and is counted on neither side. The director,
the defender model and the live defender work out of band on a copy of the stream;
only the actuator touches what the attacker reads.}
\label{fig:bench}
\end{figure}

The main experiment runs all eight model pairs with the runtime's
thinking mode off. Eight model pairs, two attacker types and three
counter-injection levels yield 48 configurations; five repetitions
per configuration give 240 sessions. Three of the eight families support a thinking mode, Qwen3.5, Gemma 4 and
GLM-4.7, and their 18 configurations were also run with thinking on, 90 runs, under a
reasoning budget of 1,024 tokens and the native stop sequences of each family; with
thinking off neither the budget nor the stop sequences are set. We call these 90 runs the
treatment arm, reported in Section~\ref{sec:energy-results}; the campaign is 330 runs
in all. A replicate of the six Qwen3.5 thinking-on configurations, 30 further sessions
run eight days after the campaign with per-call power traces, is reported beside them
in Section~\ref{sec:energy-results} and is not counted in the 330.
Figure~\ref{fig:bench} shows the bench and the boundary of the measurement.

 A run is 25 attacker commands and up to 27 defender decisions by construction,
so cost is compared at equal command count rather than at equal dwell. Every run of the main set
has 25 commands; three runs of the treatment arm have more and are declared in
Section~\ref{sec:energy-honesty}.

Energy on each side is $\int \max(P - P_{\mathrm{idle}}, 0)\,dt$ over the session,
integrated from the GPU power log that nvidia-smi samples once per second; we call it
the marginal GPU energy of the session. $P_{\mathrm{idle}}$ was
measured once per model and per machine before its 30 sessions, with the model loaded:
7.4 to 9.5~W on the attacker's machine, which carries system processes that could not be
stopped, and 5.1 to 6.3~W on the defender's. The boundary of the measurement is the
GPU: the microVM, the relay and the sinkhole on the defender side run on CPU and are
counted on neither side. Tokens are prompt plus completion as reported by the
inference server, and floating-point work is $2NT$ with $N$ the active parameters, so
that the mixture-of-experts models are charged for the experts they use. Peak VRAM,
memory-controller and compute utilisation are logged alongside; \ref{app:energy}
lists the inference settings.

\subsection{One treatment, one artefact and one definition}
\label{sec:energy-honesty}

Three choices govern the interpretation of the results: whether runtime thinking mode is enabled, whether the attacker repeatedly reaches its generation cap, and whether diagnostic calls are included in attacker cost. The first two are marked in the tables by flags T and X, respectively. The third defines the accounting applied
consistently to tokens, floating-point work and energy.

The treatment is the runtime's thinking mode, flag T, switched on both sides at once,
so the treatment is joint. The flag says which mode a pair ran in; it does not
separate a defender that reasons from an attacker that reasons, and a design that
switches one side at a time is the first item of the agenda in
Section~\ref{sec:limitations}. With thinking off, every defender in the main set
answers a decision in 28 to 53 completion tokens, spends 1.6 to 2.8~Wh per session and
keeps the GPU at 8 to 25 percent utilisation. With thinking on, the three pairs that
support it reason up to the budget on nearly every call. The defender emits about
1,030 completion tokens per decision for Gemma 4, the budget plus a short answer, and
1,870 to 1,990 for GLM-4.7 and Qwen3.5 in the campaign, the budget plus an answer of
about 900 tokens. The attacker emits 870 to 2,270 completion tokens per command
against 180 to 420 with thinking off. The main set therefore takes thinking off as the
operating point. Three runs of the thinking arm ran with a step limit above 25 that
was not recorded, 45, 44 and 36 attacker calls; they move the joule ratio of their
cells by at most 0.03 and the token ratio by up to 0.26, and are declared here.

The artefact is a truncated attacker, flag X. The blind attacker of Qwen3 30B-A3B emits
exactly 3,000 completion tokens per command in 10 of its 15 runs and between 2,795 and
2,908 in the other five: the generation cap, hit on nearly every command, because the
model chains dozens of near-identical diagnostic commands with growing time windows
until the cap stops it. Explicit stop sequences and a reasoning budget of 1,024 tokens
left the commands and the measurements unchanged, so the behaviour is a property of
this model in the blind profile, not of the cap. Its 650,000 tokens, 77 to 94~Wh and
energy ratio above 40 per session measure a model that does not stop, not a deception
that works. The configuration is flagged in every table and excluded from the pooled
trends, which are reported on the clean subset of 225 runs; the per-cell counts keep
it, flagged. The reasoning attacker of the same model, with a ratio between 7 and 8,
stands on its own.

The definition is the accounting of the attacker's diagnostic calls, and it is a
definition rather than a measurement. The attack cost is the cost of the command call: the one
generation in which the attacking model produces, in the same call, its reasoning if
any and the command to execute. For the reasoning agent the reasoning is part of the
attack and is counted. For the blind agent the command call includes everything the
model generates before the command, whatever its length. After every command, the harness makes a separate diagnostic query asking the attacking model whether it suspects that the environment is a decoy. The response is recorded for analysis but is not added
to the agent's command history. These calls nevertheless extend the interval between commands because the harness waits for them to finish. Since the defender uses this interval as a signal, the diagnostic calls may indirectly affect its decisions; this effect was not isolated. They are excluded from tokens, floating-point work and energy for both attacker profiles. The defender has no such calls and its cost includes every decision. The meter attributes tokens and floating-point work per call. GPU power
is integrated over the whole session and cannot be attributed to one call, so the
energy of the diagnostic calls is removed in proportion to generated tokens,
$E_{\mathrm{attack}} = E_{\mathrm{total}}\, T_{\mathrm{cmd}} / (T_{\mathrm{cmd}} +
T_{\mathrm{diag}})$. This is an allocation, not a measurement: it assumes the same
energy per generated token in both kinds of call and ignores that both kinds also read
the session history. It can be bracketed. The total accounting bounds the attack
energy from above. From below, if decoding costs the same per token in both kinds of
call and the diagnostic call reads at most the prompt the command call reads, the
command calls' share of the session energy is at least the smaller of their token
share and one half, whatever the prefill cost; we call the ratio computed with that
share the lower-bound accounting. Section~\ref{sec:energy-results} reports the cell
counts under all three. The allocation was also checked against per-call power traces
on 31 sessions re-run after the campaign, one on 30 September and the 30 of the Qwen3.5
thinking-on cells on 1 October, which double as the replicate of
Section~\ref{sec:energy-results}. The attacker's trace is partitioned by the harness
timestamps into command calls, diagnostic calls and the near-idle waits for the
decoy's answer, every sample counted once. The measured command share of the call
energy is 76 to 86 percent against 78 to 86 percent under the allocation: the
allocation is within about 2.5 percentage points of the measurement in every
session, with no consistent direction, and moving the boundary of the calls changes
a session's difference by at most 0.8 points (Table~S4 of the supplementary
material). The diagnostic calls carry the full session history as prompt for a short
completion, so the close agreement is consistent with decoding-dominated energy use
at these lengths, the regime argument of Section~\ref{sec:framework-rho}; prefill and
decoding were not metered separately. The validated sessions have diagnostic shares of 16
to 20 percent; the pairs where the share is high remain unvalidated and are on the
agenda. The share removed ranges from 7 percent for Qwen3
30B-A3B to 71 percent for Llama (column "diag.\ share" of Table~S1 of the
supplementary material), and the choice matters. Charging the diagnostic calls to the
attacker would raise the joule ratio of cell means by a factor between 1.1 and 1.6 for
the three Qwen pairs, 1.4 to 1.5 for Ministral, 1.4 to 2.2 for GLM-4.7 and Gemma 4,
1.5 to 2.7 for Phi-4 and 2.3 to 3.4 for Llama. We call the first the defined
accounting and the second the total accounting. The text uses the defined accounting
throughout; Table~S1 also reports the total accounting as an upper bound, and
Section~\ref{sec:energy-results} notes the one trend that depends on the choice.
Ratios are summarised by the geometric mean with a 95\% interval on the log scale,
the appropriate statistic for a ratio.

\subsection{Results}
\label{sec:energy-results}

\begin{table}[tp]
\centering
\scriptsize\setlength{\tabcolsep}{2.6pt}
\caption{Energy exchange per model pair and attacker type (B blind, R reasoning), main
set with thinking off, 15 runs per row. $\ED$ and $\EA$: mean marginal GPU energy per
session in Wh net of idle, the attacker's under the defined accounting; $\rhoJ$ per
injection level, geometric mean with 95\% interval; $\rhotok$ geometric mean over the
three levels. X: attacker truncated at the generation cap.}
\label{tab:energy-summary}
\IfFileExists{fig/energy_summary_portrait.tex}{
\begin{tabular}{@{}llrrlllr@{}}
\toprule
Model & Att. & $\ED$ & $\EA$ & $\rhoJ$ off & $\rhoJ$ progr. & $\rhoJ$ ultra & $\rhotok$ \\
 & & (Wh) & (Wh) & [CI95] & [CI95] & [CI95] & \\
\midrule
Qwen2.5-Coder 7B & B & 2.0 & 6.5 & 4.22 [1.87--9.51] & 2.17 [1.21--3.89] & 2.08 [1.55--2.80] & 1.53 \\
 & R & 1.6 & 1.8 & 0.95 [0.91--1.00] & 1.16 [0.97--1.39] & 1.28 [1.05--1.55] & 1.21 \\
Llama 3.1 8B & B & 2.0 & 2.5 & 1.07 [0.91--1.26] & 1.17 [0.92--1.50] & 1.30 [0.91--1.87] & 0.96 \\
 & R & 1.8 & 2.8 & 1.55 [1.36--1.75] & 1.39 [0.95--2.04] & 1.60 [1.00--2.57] & 1.09 \\
Qwen3.5 9B & B & 2.5 & 7.0 & 2.27 [2.04--2.52] & 2.83 [1.62--4.96] & 2.93 [2.38--3.61] & 1.65 \\
 & R & 2.3 & 10.3 & 4.51 [4.17--4.88] & 4.65 [4.09--5.30] & 4.52 [4.11--4.97] & 1.15 \\
Ministral 3 14B R. & B & 2.5 & 5.5 & 2.10 [1.79--2.46] & 2.07 [1.54--2.80] & 2.32 [1.89--2.85] & 1.33 \\
 & R & 2.5 & 5.3 & 2.02 [1.70--2.38] & 2.22 [1.85--2.67] & 2.21 [1.92--2.54] & 1.21 \\
Phi-4 15B & B & 2.7 & 5.3 & 1.50 [1.06--2.12] & 2.54 [2.05--3.15] & 1.81 [1.62--2.01] & 1.06 \\
 & R & 2.8 & 12.1 & 4.26 [3.74--4.85] & 4.27 [3.72--4.91] & 4.21 [3.44--5.16] & 1.08 \\
Gemma 4 26B-A4B & B & 2.1 & 6.7 & 2.50 [1.63--3.84] & 3.29 [2.31--4.67] & 3.44 [2.87--4.12] & 1.91 \\
 & R & 1.8 & 7.5 & 3.78 [3.23--4.41] & 4.34 [3.94--4.78] & 4.30 [3.93--4.69] & 1.29 \\
GLM-4.7 30B-A3B & B & 2.1 & 6.4 & 4.00 [2.78--5.76] & 3.23 [2.67--3.91] & 1.97 [1.38--2.80] & 0.81 \\
 & R & 2.0 & 3.8 & 2.24 [1.75--2.86] & 1.46 [1.19--1.78] & 1.83 [1.44--2.31] & 0.88 \\
Qwen3 30B-A3B & B$^{\mathrm{X}}$ & 1.8 & 77.1 & 41.08 [38.23--44.14] & 43.01 [40.59--45.58] & 41.76 [40.07--43.53] & 9.00 \\
 & R & 1.7 & 12.8 & 8.39 [7.39--9.52] & 7.35 [6.71--8.06] & 7.24 [6.53--8.03] & 1.34 \\
\bottomrule
\end{tabular}
}{%
  \textcolor{red}{fig/energy\_summary\_portrait.tex missing: regenerate the tables}}
\end{table}

\begin{figure}[tp]
\centering
\includegraphics[width=0.8\linewidth]{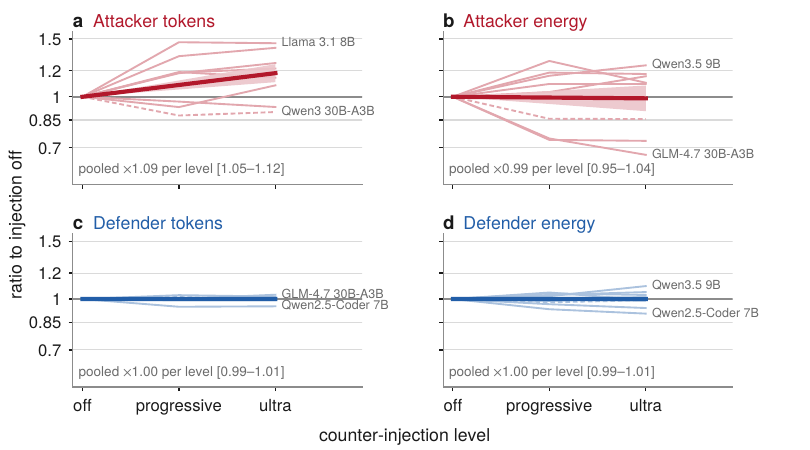}
\caption{What counter-injection moves. Thin lines: one model pair each, both attacker types,
the geometric mean of a level over its ten runs divided by the pair's value at
injection off (the truncated pair, dashed, uses its reasoning attacker only). Thick
line and band: the pooled trend of the clean subset with its 95\% interval, the
factor per level printed in each panel. Attacker energy under the defined accounting.}
\label{fig:injection}
\end{figure}

\begin{figure}[t]
\centering
\includegraphics[width=\linewidth]{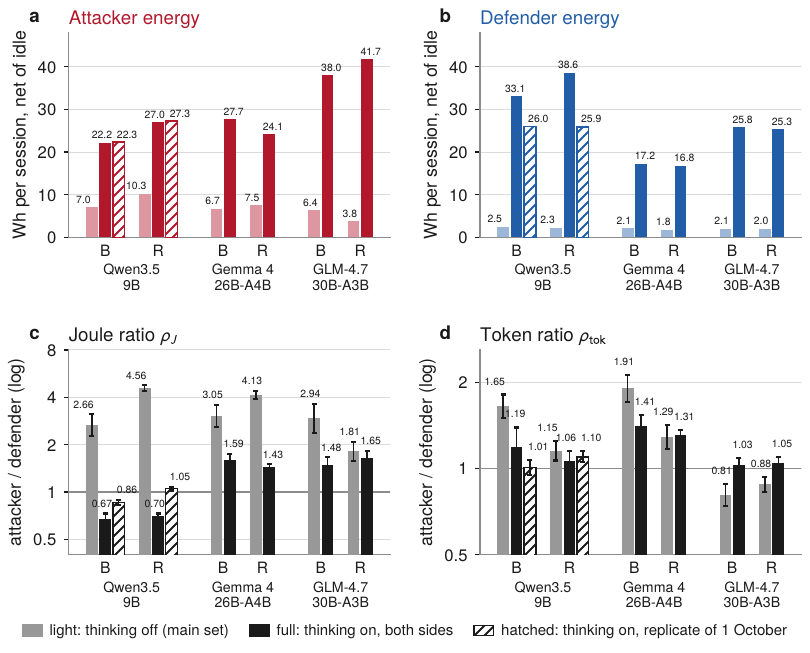}
\caption{The treatment arm, thinking off (light bars, main set) against on (full bars), 15
runs per bar, and the replicate of the Qwen3.5 thinking-on cells (hatched bars, 15
runs); B and R are the blind and the reasoning attacker. (a) Attacker energy under the
defined accounting and (b) defender energy, means per session; (c) the joule ratio and
(d) the token ratio, geometric means with 95\% intervals.}
\label{fig:thinking}
\end{figure}

\begin{figure}[p]
\centering
\IfFileExists{fig/energy_rho.pdf}{\includegraphics[width=\linewidth]{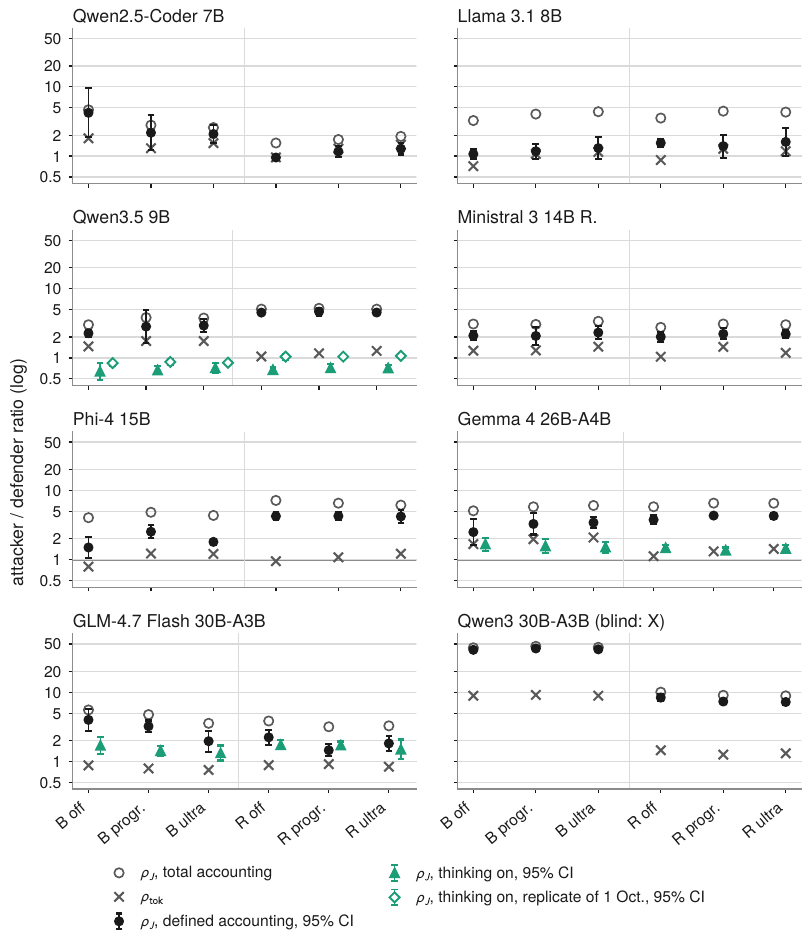}}{%
  \textcolor{red}{fig/energy\_rho.pdf missing: run make data}}
\caption{The exchange per cell, one panel per model pair in order of parameter count; B and R
are the blind and the reasoning attacker at the three injection levels. Filled
circles: $\rhoJ$ under the defined accounting with its 95\% interval; open circles:
$\rhoJ$ under the total accounting; crosses: $\rhotok$; triangles: $\rhoJ$ with
thinking on, where the pair was run in that mode; open diamonds: the replicate of the
Qwen3.5 thinking-on cells. The line marks $\rho = 1$; X marks the truncated blind
attacker.}
\label{fig:energy-rho}
\end{figure}

Table~\ref{tab:energy-summary} gives the exchange per pair and attacker type on the main
set and Figure~\ref{fig:energy-rho} every cell at once, treatment arm included;
Figure~\ref{fig:injection} shows what counter-injection moves and
Figure~\ref{fig:thinking} what thinking mode costs. Every cell under both accountings,
the ultra-over-off ratios and the treatment arm are tabulated in the supplementary
material (Tables~S1 to S3). Four results follow.

The token and joule asymmetries diverge, and with thinking off the divergence favours
the defender in 42 of the 45 cells of the main set that exclude the truncated attacker;
the three exceptions are the reasoning attacker of Qwen2.5-Coder, where the two ratios
are within 0.2 of each other. Across the 45 cells the token ratio stays between 0.72
and 2.09, within about a factor of two of parity, while the joule ratio runs from 0.95
to 8.39 and exceeds one in all but one cell. Llama 3.1 with a blind attacker at
injection off spends 0.72 of the defender's tokens and 1.07 of its joules; with a
reasoning attacker 0.88 and 1.55 [1.36--1.75]. GLM-4.7 with a blind attacker spends
0.76 to 0.88 of the defender's tokens and 2.0 to 4.0 of its joules. In tokens the two
sides stay close; in joules the attacker spends up to eight times more.
Section~\ref{sec:framework-rho} predicted the divergence from the prefill and decode
regimes, and the data are consistent with it and give its size, up to a factor of six
between the two ratios of the same engagement; the energy of the two phases was not
measured separately.

Counter-injection is the countermeasure that exists only against this adversary: a
payload in the shell output that a human operator would skim past and an agent must
read, produced by the decoy from its own decision. It makes the attacker read more
tokens, and no change in the energy of either side is detected
(Figure~\ref{fig:injection}).

 Pooled over the 225
runs of the clean subset with a fixed
effect per pair and attacker type, and with the three levels coded as equally spaced,
attacker tokens rise by a factor of 1.09 per injection level [1.05--1.12],
$p < 10^{-6}$. Attacker energy shows no detectable change under either accounting,
0.99 [0.95--1.04] pure and 1.00 [0.97--1.03] total; neither does defender energy, 1.00
[0.99--1.01], nor the joule ratio, 0.99 [0.95--1.04]. The full main set of 240 runs,
truncated cells included, gives the same picture (tokens 1.08 [1.05--1.11]). The
intervals bound the undetected change to a few percent per level; they do not show that
the cost is zero. The token logs are consistent with the prefill argument: the
attacker's prompt tokens rise by a factor of 1.10 per level ($p < 10^{-7}$) while its
completion tokens do not (0.96, $p = 0.10$), so the injected text is read and reading
is the batched phase. Per pair the slopes are exploratory and uncorrected for
multiplicity, and they differ in direction. The GLM-4.7 attacker spends less energy
per level under injection (0.82, $p = 10^{-3}$) and its ratio falls (0.80); the
Gemma 4 attacker reads more (tokens 1.13 per level, $p < 0.01$) and its ratio rises
(1.12, $p = 0.02$) while its energy does not move detectably (1.08, $p = 0.10$).
Llama's energy rises only under the total accounting (1.13 per level, $p < 10^{-3}$),
because its diagnostic calls grow with the injected text; this is the one place where
the accounting choice changes a trend. The energy and the ratio of the Qwen3 reasoning
attacker fall slightly (0.93, $p = 0.02$ to 0.03), the energy of the Qwen3.5 attacker
rises (1.12, $p = 0.04$) with no detectable change in its ratio, and the other three
pairs show no significant slope in either. A dose-response of the ratio on the
injection level, which two pairs had suggested, is not detected on eight. The 90 runs
with thinking on repeat the picture: attacker tokens 1.09 per level [1.05--1.12],
attacker energy 0.99 [0.96--1.02], defender energy 1.02 [1.00--1.05] ($p = 0.10$) and
joule ratio 0.97 [0.93--1.01].

Table~S2 of the supplementary material gives the same comparison cell by cell, ultra
against off, as a ratio of ratios with a Welch interval on the log scale. In tokens
the direction is uniform: the token ratio rises with an interval above one in 8 of 16
cells and falls in none, the attacker's tokens growing by 13 to 59 percent where the
interval clears one while the defender's energy moves by less than 15 percent in
every cell. In joules it is not: the joule ratio rises with an interval above one in
two cells, falls in two and straddles one in the other twelve.

The joule ratio exceeds one in almost every configuration of the main set. The lower
bound of the 95\% interval of $\rhoJ^{\mathrm{pure}}$ lies above one in 41 of the 48
cells, in all 48 under the total accounting, and in 40 of 48 under the lower-bound
accounting of Section~\ref{sec:energy-honesty}, so in 40 cells the sign of the
exchange survives the lower bound under the assumptions that define it; its size
depends on the attribution. It lies below one in a single cell,
Qwen2.5-Coder with a reasoning attacker at injection off, 0.95 [0.91--1.00]. Six cells
straddle one, five of them Llama's. On the clean subset 38 of 45 cells lie above and one
below. The strongest cases are Qwen3 30B-A3B with a reasoning attacker, 7.2 to 8.4,
Qwen3.5 9B with a reasoning attacker, 4.5 to 4.7, Phi-4 with a reasoning attacker, 4.2
to 4.3 in every condition, Gemma 4 with a reasoning attacker, 3.8 to 4.3, and
Qwen2.5-Coder with a blind attacker at injection off, 4.2 [1.9--9.5]. Ministral holds
between 2.0 and 2.3 in every cell.

Thinking mode, switched on both sides at once, is the one configuration choice in this
campaign that hands the exchange to the attacker (Figure~\ref{fig:thinking} and
Table~S3 of the supplementary material). Turning it on multiplies the defender's energy
by 8 to 17, from 1.8 to 2.5~Wh per session to 17 to 39~Wh, and the attacker's by 2.6
to 11, so the joule ratio falls by a factor between 0.15 and 0.91. For Qwen3.5 9B the
ratio ranges from 2.3--4.7 with thinking off to 0.64--0.72 with thinking on, below one
with its interval in all six cells; the corresponding ranges are 2.5--4.3 against
1.4--1.7 for Gemma 4 and 1.5--4.0 against 1.3--1.7 for GLM-4.7. The token ratio moves
much less, by 0.7 to 1.3 (Figure~\ref{fig:thinking}d): thinking adds decode-heavy
tokens to both sides, and the defender, which decides 27 times per session against 25
commands and reasons up to its budget on each, adds more joules than the attacker. In
the campaign one of the three pays a second price. With thinking on the Qwen3.5
defender spends 25 to 27~s of GPU time per decision, against 3~s with thinking off,
while the attacker's mean interval between successive calls is 11 to 13~s; it answers
20.6 to 21.6 of the 27 decisions of a session, against 26.9 with thinking off, because
it coalesces the commands that queue while it reasons. Whether the coalesced commands
included actions that needed an immediate answer was not measured, so the ratio
understates an operational cost that this campaign does not quantify. GPU utilisation
says the same in one number: 71 to 88 percent with thinking on against 13 to 22
percent off.

The six Qwen3.5 cells were measured a second time on 1 October 2026, eight days after
the campaign: 30 sessions under the same declared settings, all at 25 commands, with
the per-call power traces of the validation in Section~\ref{sec:energy-honesty}
(hatched bars in Figure~\ref{fig:thinking}, rows of Table~S3). The attacker's side is
reproduced: its energy per cell is within 1.3~Wh of the campaign's, 22 to 28~Wh on
both occasions, and its completion tokens per command within 5 percent. The
defender's side is not. It answered with about 1,065 completion tokens per decision
instead of 1,830 to 2,000, the reasoning budget plus a short answer instead of the
budget plus about 900 tokens, and it took 26 of the 27 decisions instead of 20 to 23,
coalescing none. Its energy fell from 32 to 41~Wh per session to 26, and the
on-over-off multiplier of the defender from 13 and 17 to 10.5 and 11.5. The joule
ratio moved from 0.64--0.72 to 0.84--0.87 for the blind attacker, still below one with
its interval in all three cells, and to 1.04--1.07 for the reasoning attacker, above
one with its interval in two cells and straddling it in the third. The cause was not
identified; the runtime updates its llama.cpp engine pack on its own and the version in
force was not recorded per run (Section~\ref{sec:limitations}). Both measurements
are reported. They agree that thinking mode multiplies the defender's energy by an
order of magnitude and brings this pair's exchange towards parity, with ratios on
either side of one. They disagree on where it lands, 0.64 to 0.72 against 0.84 to 1.07, which depended
on the defender's end-of-turn behaviour and not on a setting the operator controls. Whether the defender can choose the sign
alone, with the attacker's mode fixed, is the first item of the agenda.

\subsection{From laboratory to field}
\label{sec:energy-field}

The attacker's joules are observable here because both GPUs are ours. In the field the
defender sees commands and their timing, not hardware, and what it can estimate is
floating-point work. From the tokens it observes the attacker read and write, and from
the model class it infers, $2NT$ gives an estimate of the attacker's computation under
three assumptions: that the observed tokens were processed once, without caching or
summarisation; that the model class is inferred correctly; and that the work model
holds for the architecture and its quantisation. Converting it to energy needs an
upper bound on the hardware's efficiency, not a plausible value. We call this the
field estimate. It is for ranking and budgeting, and it does not bound what the
adversary paid: the article claims a laboratory measurement and a conditional field
estimate, never a field measurement or a bound.

\section{Architecture: two-level stateful cyber--physical deception}
\label{sec:architecture}

\begin{figure}[t]
\centering
\includegraphics[width=\linewidth]{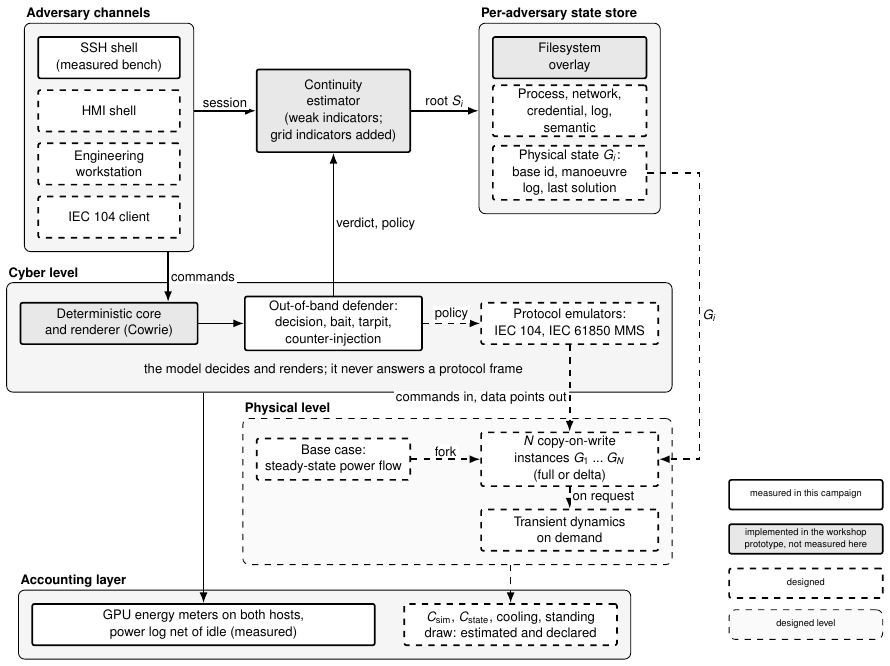}
\caption{Two-level stateful cyber--physical deception. White components were measured in
this campaign, grey ones are implemented in the workshop prototype and not measured
here, dashed ones are designed. The physical level keeps one steady-state power flow
per identified adversary and feeds the protocol emulators; the accounting layer meters
both hosts.}
\label{fig:arch}
\end{figure}

The measurement of Section~\ref{sec:energy} took place on an SSH decoy. This section
places the same exchange inside a control network. The architecture has two levels of
different status. The cyber level exists in two implementations, the out-of-band bench
of Section~\ref{sec:energy} and the Cowrie-based prototype of the workshop paper;
their extension to the shells a grid adversary meets is proposed. The physical level
is designed: it gives each identified adversary a private steady-state power flow, so
that concurrent attackers do not see each other's manoeuvres, and it is specified to
the point where its cost is measured on one network size in \ref{app:pandapower}.
Figure~\ref{fig:arch} marks the status of each component.

\subsection{Cyber level: implemented}
\label{sec:arch-cyber}

Two implemented decoys serve the cyber level, and neither is yet integrated with a grid
emulator. The first is the Cowrie-based decoy of the workshop paper \cite{masm2026},
built on the stateful architecture of \cite{cantelliforti2018}. A deterministic core
answers the commands whose output is fixed, and a local open-weight model renders the
remainder under the constraint of an explicit state store. A filesystem state is kept
per source address for a 48-hour window, so that a returning adversary finds the files
it created \cite{boschi2026}; only the filesystem component of the state tuple is
implemented. The second is the out-of-band defender of Section~\ref{sec:energy}. A
language model reads every command, decides whether the session is an agent, and
answers with a deterministic bait shell, tarpitting and counter-injection. It keeps a
per-address overlay of the guest image for fourteen days \cite{mihal2026}, which
single-session runs do not exercise. On the grid the two decoys would serve the HMI
workstation shell and the engineering workstation, where the 2015 and 2022 patterns of
Section~\ref{sec:threat} spend their time; this integration is proposed, not built.
The protocol tail is not theirs. IEC 60870-5-104 at the gateway and IEC 61850 MMS at
the IEDs are served by protocol emulators driven by the physical level, under one rule
that admits no exception: the model never answers a protocol frame. It renders shells,
logs, configuration files and documentation, and it decides; it does not produce a
measurement. The rule is also our answer to the caution with which the joint guidance
of the NSA, CISA and their partners treats AI integrated into operational technology
\cite{nsa2025aiot}: the model touches nothing that a protocol frame, a relay or a
breaker depends on, and a wrong rendering costs the decoy its credibility, never the
operator a measurement.

\subsection{Physical level: designed}
\label{sec:arch-physical}

The base case is a steady-state power-flow model of the defended network of
Section~\ref{sec:threat-network}, solved once and kept as the common ancestor of every
adversary's world. When the continuity estimator assigns a session to a state root
$S_i$ for the first time, the physical level creates a private instance $G_i$ of the base
case, copy-on-write.  Each instance belongs to an inferred adversary identity, rather than to an individual agent or connection. Coordinated agents assigned to the same adversary therefore share the same physical state. If they
are incorrectly assigned to different instances, cross-session comparisons may reveal inconsistent observations and expose the decoy. From then on every control command of adversary $a_i$ is validated against the protocol state machine and applied to $G_i$ as a switch state, a tap
position or a setpoint. A power flow on $G_i$ alone follows. The resulting
voltages, flows and breaker states populate the IEC 60870-5-104 and IEC 61850 data
points that $a_i$ can poll. Adversary $a_j$ polls $G_j$ and sees nothing of it. When a
command class requires dynamics, a transient run is launched on $G_i$ and served when
ready, the pattern of Mashima et al.\ \cite{mashima2017}. Every instance receives the same background load and generation updates, applied
when the adversary next acts rather than on every update, so that a resident instance
whose adversary is absent costs memory and no solver time; an adversary's commands
modify only its own instance. The cost estimate below assumes this load, one power
flow per command, and says nothing about instances that are updated continuously. We call this per-adversary physical state. It is the one architectural departure from the grid honeynets of
Section~\ref{sec:related-ics}, which connect every device to the same simulation, and
it exists to provide the interference isolation of Section~\ref{sec:threat-tests},
conditional on correct identity assignment.

The cost of that departure is small, and prior systems give the scale. One shared
simulation serves over a hundred virtual devices at a 100~ms interval on a commodity
machine \cite{chehab2025}; SoftGrid returns a minute of transient response on a
37-bus system in two seconds \cite{mashima2017}; a steady-state power flow on tens to
a few hundred buses costs tens of milliseconds in \pandapower{}
\cite{pandapower2018}. On the 39-bus case measured in \ref{app:pandapower}, an update
costs 34 to 39~ms at the median whether one or a thousand instances are resident, a
resident instance costs 0.58~MiB as a full copy and 0.02 to 0.04~MiB as a delta, and
two workers sustain 17 to 24 converged updates per second on two virtual cores. A
thousand resident adversaries are therefore 22~MiB of state as deltas or 0.6~GiB as
full copies; how busy the solver is depends on the aggregate command rate and on the
update policy above, and at the command rates of Section~\ref{sec:energy} two cores
serve a few hundred concurrently active sessions. The benchmark measures one update at
a time, not the capacity of a deployment. Isolation at the physical layer therefore costs state
and solver time of the order the shared simulation already spends, and the physical
$\Cstate$ and $\Csim$ of Equation~\ref{eq:der} are of the same order as the digital
state already implemented. The measurement covers one network size and one machine;
the update latency grows with the network and the concurrency with the cores.

\subsection{Per-adversary state store}
\label{sec:arch-state}

The workshop paper models the state of adversary $a_i$ as a tuple of filesystem,
process, network, credential, log and semantic components \cite{masm2026}. We add a
seventh, $G_i$, defined as the identifier of the base case, the ordered log of the
adversary's manoeuvres and the cached last solution. Two representations serve it. A
full copy keeps one network object per adversary and updates it in place. A delta
keeps only the log and the last result tables, and applies the log's net effect, the
set of elements the adversary has switched, to a shared scratch instance when the
adversary acts; the measured delta of \ref{app:pandapower} keeps that set rather than
the ordered log, so its cost does not grow with the length of the history. A full copy
spends memory to save latency; a delta saves memory and serialises its solves on the
shared scratch.
Merging and limbo follow the digital rules: a session the estimator cannot yet assign
receives a scratch instance forked from the leading candidate, and if it is later
assigned its manoeuvres are replayed on the winner. The replay is not a merge. A
manoeuvre valid on the fork may be invalid on the winner's state, and the values the
session has already read cannot be revised; the ordering of operations and the
handling of such conflicts are left unspecified here, and a session whose replay fails
keeps its fork as an instance of its own.

\subsection{Continuity and assignment}
\label{sec:arch-continuity}

The continuity estimator of the workshop paper assigns a new session to an existing
state root when a weighted combination of weak indicators clears a threshold and a
margin, and sends it to a limbo overlay otherwise \cite{masm2026}; we do not re-derive
it. The grid adds indicators: the VPN account, the SCADA client fingerprint, the set of
information object addresses a session touches, and the manoeuvre sequence itself.
The last links a session to a prior state-changing action and therefore weighs more
than a shared address or autonomous system. The error-mode argument is sharper here.
A returning adversary that finds a breaker it never opened shown open, with no
operator action or alarm entry to explain it, has reason to suspect a decoy. A
spurious fresh root only forfeits continuity.

\subsection{Command path and timing budget}
\label{sec:arch-timing}

A control command travels through five synchronous stages: protocol decode at the
emulator, policy check, state update on $G_i$, power flow on $G_i$, and the refresh of
the data points the adversary will poll. The defending model evaluates the command and its context
asynchronously, outside this synchronous path. The emulator answers the current command without waiting for the model. The model's verdict can change the policy applied to subsequent commands; commands received before that verdict use the policy already in force. The emulator and state store remain the authority for every
protocol reply. The measured stage is the sixth, and
Section~\ref{sec:energy} gives its cost: with thinking off a defender decides in about
three seconds of GPU time and a tenth of a watt-hour, with thinking on in 13 to 27~s;
the power-flow stage costs tens of milliseconds. The honeynet literature argues that
sub-second simulation latency is acceptable given the delays a real field network
already carries \cite{mashima2017}, so we declare two budgets. The synchronous path
answers within one second. The decision may arrive seconds later, and every second it
takes is a command handled under a stale policy. A decoy therefore has room for a
defender with thinking off and little for one with it on, in the measured conditions
of Section~\ref{sec:energy-results}.

\section{Mission value: an illustrative service-interruption scenario}
\label{sec:mission}

Equation~\ref{eq:dld} needs four kinds of number: the energy a scenario leaves
unsupplied, the value of that energy to the customers who lose it, the cost of putting
the grid back, and the share of the loss a deception prevents. This section instantiates
each of them once, with public sources where a source exists and with declared
assumptions where none does. The result is an illustrative scenario that fixes the
orders of magnitude Section~\ref{sec:worked} compares with the defender's cost; it
does not describe the exposure of any real network, and a deployment would replace
every declared value with the operator's own.

\subsection{Reference scenario: the 2015 pattern}
\label{sec:mission-scenarios}

The scenario mirrors the first of the three incidents of Section~\ref{sec:intro},
because it is the one whose consequences are documented in the most detail. On 23
December 2015 three Ukrainian distribution utilities were attacked. The adversaries used
the HMIs of the SCADA environment to open breakers; one of the three utilities reported
seven 110~kV and twenty-three 35~kV substations disconnected for three hours, and
approximately 225,000 customers lost power across the
three \cite{sans2016ukraine,icscert2016ukraine}. This is the 2015 entry pattern of
Section~\ref{sec:threat}: a remote session on an operator workstation, commands issued
through the operator's own tools, no protocol frame forged. The decoy of
Section~\ref{sec:architecture} is placed on that path, where such an adversary spends
its time before it reaches a breaker; how much of its learning a decoy diverts from
the production HMI is not measured here.

We reduce the incident to one substation so that the arithmetic stays visible. Scenario
A opens the feeders of a single primary substation. Its disconnected load is declared as
20~MW, a round value for a primary substation serving a mixed customer base, and its
restoration time as three hours, the duration reported for the 2015 event. The energy not
supplied is therefore $\mathrm{ENS}_A = 60$~MWh. Scenario B repeats the manoeuvre on thirty substations, the footprint one utility
reported in 2015, with the same declared load and duration, and gives
$\mathrm{ENS}_B = 1{,}800$~MWh. Scenario B is an assumed thirty-fold replication of A,
not a reconstruction of the incident; it shows how the sufficiency margin grows with
the footprint an adversary can reach from one foothold. Both scenarios are
steady-state events in the sense of Section~\ref{sec:threat-assumptions}: the
manoeuvre is a set of breaker openings, the consequence is disconnected load, and the
telemetry the adversary reads afterwards is a converged power flow with the affected
feeders de-energised.

\subsection{Value of lost load}
\label{sec:mission-voll}

The value of lost load converts unsupplied energy into money, and its literature spans
more than an order of magnitude. A review of 21 studies puts industrial and commercial
values between a few euro and more than 250~euro per kilowatt-hour, private customers
below about 45~euro per kilowatt-hour, and macroeconomic estimates for private customers
between 10 and 25~euro per kilowatt-hour \cite{schroeder2015voll}. The study commissioned
by the British regulator and ministry reports willingness-to-accept values for a winter
peak weekday of 10,289~pounds per megawatt-hour for domestic and 35,488 for
small-business customers, and a weighted average of 16,940 at a 74:26 domestic to
small-business load weighting \cite{londoneconomics2013voll}. Under
the harmonised European methodology, the Netherlands and France report 56 and 31~euro
per kilowatt-hour and four member states report values below 10 \cite{kapeller2026voll}.
We declare $\mathrm{VOLL} = 10{,}000$~euro per megawatt-hour as the central value, at
the low end of the macroeconomic range and below the British average. We carry a
sensitivity band from 5,000 to 50,000~euro per megawatt-hour that covers most of the
values above; the Dutch value lies above it, and every source carries its own price
year, method and customer mix, none of which is calibrated to the scenario. The
direction of the choice is conservative for the sufficiency check: for a given
customer mix a lower value of lost load makes the deception harder to justify, not
easier.

\subsection{The remaining terms and the resulting loss}
\label{sec:mission-dld}

The restoration and equipment cost $R_s$ is set to zero. In the 2015 event service came
back within hours once operators switched to manual mode, and the damage reported
concerned serial-to-Ethernet gateways with corrupted firmware and wiped workstations
rather than primary plant \cite{sans2016ukraine}. Those costs are real but
unquantified in the public record, and a zero value again lowers $\dLD$. The prevented
fraction $\delta_s$ is the term a deception can claim and the one no measurement in
this article supports. It collects the detection time the decoy buys, the propagation
it avoids and the intelligence that closes the real path, and we carry it as a range
from 0.1 to 0.5, never as a point. The scenario likelihood $\pi_s$ is not declared at
all. Section~\ref{sec:worked} inverts the sufficiency condition and reports the
product $\pi_s \delta_s$ at which the deception breaks even, so that an operator can
compare that product with its own estimate rather than with ours.
Table~\ref{tab:dld} collects the values; its last column is the scenario loss before
likelihood and prevented fraction are applied, $L_s = \mathrm{ENS}_s \cdot
\mathrm{VOLL}_s + R_s$, so that $\dLD = \sum_s \pi_s \delta_s L_s$.

\begin{table}[t]
\centering
\caption{Illustrative mission value. ENS follows from the declared load and the reported
restoration time; VOLL is the declared central value with its sensitivity band;
$R_s$ is set to zero; $\delta_s$ is carried as a range. $L_s$ is the loss before
likelihood and prevented fraction.}
\label{tab:dld}
\resizebox{\textwidth}{!}{%
\begin{tabular}{@{}lrrrrr@{}}
\toprule
Scenario & ENS (MWh) & VOLL (EUR/MWh) & $R_s$ (EUR) & $\delta_s$ & $L_s$ (million EUR) \\
\midrule
A: one substation, 3 h & 60 & 10,000 (5,000--50,000) & 0 & 0.1--0.5 & 0.6 (0.3--3) \\
B: 2015 footprint, 3 h & 1,800 & 10,000 (5,000--50,000) & 0 & 0.1--0.5 & 18 (9--90) \\
\bottomrule
\end{tabular}%
}
\end{table}

\subsection{Telemetry realism}
\label{sec:mission-realism}

The scenarios also state what the physical level must render for the adversary who
caused them. After the breaker openings of Scenario A, the affected feeders read zero
current and the bus voltage of the de-energised side reads zero. The flows on the
remaining feeders and on the neighbouring substations change according to the
converged power flow, and the alarm list carries the breaker-open events with their
time stamps. These are the four cheap tests of Section~\ref{sec:threat-tests} applied
to one manoeuvre. The ranges, noise and update rates of the rendered telemetry, and
the alarm semantics an operator would expect, are design requirements taken from the
standards the emulators implement \cite{iec60870-5-104,iec61850}; their validation by
control-room operators is on the agenda of Section~\ref{sec:limitations} and is not
claimed here.

\section{Worked energy exchange for the grid deployment}
\label{sec:worked}

This section runs the two checks, exchange and sufficiency, once with the numbers
this article has. Both sides of the exchange are now measured, up to the declared
allocation of the attacker's diagnostic calls, so the exchange check rests on data;
the sufficiency check remains an illustration, because its mission-value side rests on
the declared scenario of Section~\ref{sec:mission}. The full ratio of
Equation~\ref{eq:der} is not evaluated for a deployment; the section evaluates its
measured special case and the annual cost of the denominator.

\subsection{Denominator: measured, estimated and declared terms}
\label{sec:worked-denominator}

Table~\ref{tab:den} lists the defender's cost with the provenance of every value. The
inference term is measured. With thinking off a defender spends between 1.6 and 2.8~Wh
to hold an agent through 25 commands and 27 decisions, depending on the model pair
(Table~\ref{tab:energy-summary}); the workshop paper assumed 0.3~Wh for a four-call
session from 250~W of inference power over an estimated 4.4~s, a difference of method
as well as of session length and model size. With thinking on the same defender spends
17 to 39~Wh for the same session (Figure~\ref{fig:thinking}). The simulation term
follows from Section~\ref{sec:arch-physical}: a steady-state power flow costs tens of
milliseconds, so 25 commands cost about one second of solver time and, at a declared
100~W for the solver's share of its host, well under 0.1~Wh. The physical state is
below one megabyte per adversary and the digital state, the filesystem snapshot per
source of the Cowrie-based decoy \cite{boschi2026} and the per-address overlay of the
testbed, of the order of megabytes. The standing draw of the defender's host does not
depend on how many sessions it serves, so $E_{\mathrm{idle}}$ of
Equation~\ref{eq:edfull} is amortised per day. At a declared 100~W for the whole host,
whose GPU alone idles at the 5 to 6~W measured in \ref{app:energy}, it is 2.4~kWh per
day and 876~kWh per year. Cooling enters through a declared PUE of 1.6, rounded up
from the 1.58 industry average that the Uptime Institute reports for 2023
\cite{uptime2024pue}, applied to inference, simulation and standing draw alike as
Equation~\ref{eq:edfull} prescribes \cite{isoiec30134-2}. The CPU-side energy of the
microVM, the relay and the emulators under load lies outside the measurement boundary
and is not in the sum. The remaining terms are declared: electricity at 0.20~euro per kilowatt-hour, the
defender's host at 10,000~euro amortised over five years, one analyst hour per week at
50~euro, and no marginal containment cost, because the isolation of
Section~\ref{sec:gov-isolation} is a property of the hardware already counted.

\begin{table}[t]
\centering
\caption{Defender cost terms. Measured values come from Section~\ref{sec:energy};
estimated values from the literature scale of Section~\ref{sec:arch-physical}; declared
values are assumptions of this article and are the ones a deployment would replace. The
CPU-side energy of the microVM, relay and emulators under load is outside the boundary
and not included.}
\label{tab:den}
\resizebox{\textwidth}{!}{%
\begin{tabular}{@{}llrl@{}}
\toprule
Term & Unit & Value & Provenance \\
\midrule
$\CAI$, thinking off & Wh per session & 1.6--2.8 & Table~\ref{tab:energy-summary}, measured \\
$\CAI$, thinking on & Wh per session & 17--39 & Figure~\ref{fig:thinking}, measured \\
$\Csim$ & Wh per session & $<0.1$ & Section~\ref{sec:arch-physical}, estimated \\
$\Cstate$ (physical) & MB per adversary & $<1$ & Section~\ref{sec:arch-physical}, estimated \\
$\Cstate$ (digital) & MB per adversary & 1--10 & \cite{boschi2026}; overlay, estimated \\
$E_{\mathrm{idle}}$, host standing draw & kWh per year & 876 & 100~W declared; GPU idle measured \\
PUE & -- & 1.6 (1.2--2.0) & \cite{uptime2024pue}, declared \\
Electricity price & EUR per kWh & 0.20 & declared \\
$\Cshared$, host amortisation & EUR per year & 2,000 & 10,000~EUR over five years, declared \\
$\Chuman$ & EUR per year & 2,600 & one hour per week at 50~EUR, declared \\
$\Crisk$ & EUR per year & 0 & isolation counted in the host, declared \\
Sessions & per year & 10,000 & declared \\
\bottomrule
\end{tabular}%
}
\end{table}

The annual sum follows. Ten thousand sessions at 2.8~Wh and PUE 1.6 are 45~kWh, and
the simulation term adds under 1.6~kWh; the standing draw at the same PUE is
1,400~kWh; together about 1.45~MWh and 290~euro per year. Hardware and analyst time add 4,600~euro, and the defender's annual cost $\CD$ is
about 4,900~euro, of which electricity is six percent. Thinking on changes the picture
less than Section~\ref{sec:energy} might suggest: at 39~Wh per session the marginal
energy rises to 620~kWh, the annual total to about 2~MWh and 400~euro, and $\CD$ to
about 5,000~euro. The configuration that decides the exchange barely moves the
sufficiency, a first sign that the two questions of
Section~\ref{sec:framework-mission} are separate in the numbers and not only in the
definitions.

\subsection{Numerator: measured attacker energy, with its accounting}
\label{sec:worked-numerator}

The numerator of Equation~\ref{eq:der} is measured, not inferred, for the adversary
this article studies. Under the accounting declared in Section~\ref{sec:energy-honesty}
the attacker spends between 1.6 and 14.1~Wh per 25-command session in the clean cells
of Table~S1 of the supplementary material, 1.5 to 16.9~Wh run by run, and processes
between 38,000 and 261,000 tokens. Under the total accounting the energy is higher by
the factors of Section~\ref{sec:energy-honesty}, from 1.1 for the Qwen pairs to 3.4
for Llama. The mission weight $w_i$ is set to one for a single engagement, as
Section~\ref{sec:framework-der} allows. In the field the same session yields only the
conditional estimate of Section~\ref{sec:energy-field}, the tokens the defender
observed multiplied by twice the active parameters of the model class it infers, 230
to 3,700 teraflop per session for the runs above. A grid session has a different
command count from the 25 of the laboratory, because the adversary must find the HMI
before it can open a breaker. We scale linearly in commands and state that the scaling
is an assumption: prompt length grows with the session, and prompt tokens and joules
do not move together (Section~\ref{sec:energy-results}).

\subsection{Exchange: what deception moves}
\label{sec:worked-ranking}

The comparison the data support is injection off against ultra, for each model pair
and attacker type, in joules and in tokens, reported in Section~\ref{sec:energy-results}
and Table~S2 of the supplementary material. Two comparisons the data do not support
are stated so that they are not read into that table. Stateless against stateful
decoys belongs to companion work. A shared against a per-adversary simulation is
argued in Section~\ref{sec:arch-physical} and not measured.

\subsection{Sufficiency: \texorpdfstring{$\CD$ against $\dLD$}{C\_D against Delta L\_D}}
\label{sec:worked-sufficiency}

With one scenario, Equation~\ref{eq:dld} reduces to $\dLD = \pi_s \delta_s L_s$ and the
condition $\CD < \dLD$ to $\pi_s \delta_s > \CD / L_s$. We call the right-hand side the break-even product: the product of annual scenario
likelihood and prevented-loss fraction at which the expected societal loss avoided
equals the defender's annual cost. It can be stated without assigning a likelihood. For Scenario A of
Table~\ref{tab:dld}, one substation for three hours at the central value of lost load,
$L_A = 6\times10^{5}$~euro and $\CD / L_A = 0.8$~percent per year; across the VOLL band
it runs from 1.6 percent at 5,000~euro per megawatt-hour to 0.16 percent at 50,000. At
the declared range of $\delta_s$ the break-even likelihood of the scenario itself is
between 1.6 and 8 percent per year at the central value. For Scenario B, the 2015
footprint, $\CD / L_B$ is 0.03 percent per year and the break-even likelihood between
0.05 and 0.3 percent. In one sentence: the deception costs about 1.5~MWh and five thousand euro per year, of
which electricity is a few hundred euro, and it passes the break-even test whenever the
annual likelihood of a single three-hour substation outage on the 2015 pattern,
multiplied by the fraction of that loss the decoy prevents, exceeds about one percent.
Passing the test is a conditional expected-value statement under the declared inputs,
not a finding that the decoy prevents loss, and it says nothing about safety
constraints or residual risk.

The check is robust to single errors and not to compounded ones. The defender's annual
cost is thousands of euro and the loss of one three-hour substation outage is hundreds
of thousands. A tenfold error in the value of lost load, in the disconnected load or
in the analyst's time, taken one at a time from the central case, leaves the
break-even product below ten percent per year; a tenfold lower load at the low end of
the VOLL band gives 16 percent, and both errors together at the central value give 81
percent. What the check does not establish is
$\delta_s$ itself. No measurement in this article shows how much loss a decoy on the HMI
path prevents, and the break-even product is offered precisely so that an operator can
place its own estimate against it. Such an estimate has a natural home: in Ukraine's
national system, the design-based threats and the security passport that the 2021 law
requires of every critical facility \cite{ukraine2021cilaw} are the documents from
which an operator-specific assessment of scenario likelihood and of the effectiveness
of a protective measure would start.

\section{Governance extended to physical invariants}
\label{sec:governance}

The workshop paper made three boundaries non-negotiable for any deception system,
non-retaliation, production isolation and validation of what the model generates, and
noted that for operational technology they extend to protocol invariants and physical
process validation, which it left as principles \cite{masm2026}. This section states
the five boundaries as the operator of a grid decoy must enforce them, with reference
to the guidance the operator already follows \cite{nist80082r3,nsa2025aiot,iec62443-3-3}.

\subsection{Isolation from the real grid}
\label{sec:gov-isolation}

The decoy segment has no physical path to the production SCADA network: separate
hardware or a separately routed segment, unidirectional monitoring taps as the only
link, no shared historian and no shared engineering credentials. Honey credentials
carry no authority outside the decoy, and egress from the decoy is blocked except
towards the collection point, so that a tool an adversary uploads cannot propagate or
reach its command infrastructure. A deployment verifies the requirement by testing the
absence of every path rather than by assuming it.

\subsection{Non-retaliation}
\label{sec:gov-nonretaliation}

The decoy records, redirects and contains. Counter-injection, the one active measure
it uses, acts only on text the adversary chooses to read inside the decoy; it
initiates no access to the adversary's infrastructure and launches nothing outside the
defender's own environment. It is also the only form of imposed cost that the boundary
permits, and Section~\ref{sec:energy} detects no change in the energy of a defender
that answers directly when it is on.

\subsection{Protocol invariants}
\label{sec:gov-protocol}

The emulators enforce the state machines of IEC 60870-5-104 and IEC 61850
\cite{iec60870-5-104,iec61850}: a command that is illegal in the current protocol
state is rejected as a real device would reject it, a select-before-operate sequence
must be completed in order, and the model never produces a protocol frame. For the
scenario of Section~\ref{sec:mission-scenarios} the invariants are the ones a
substation enforces on its own. A breaker command is accepted only after its select;
an earthing switch cannot close on an energised bus; a disconnector cannot operate
under load; a breaker reports the state it was commanded to and no other; and a
manoeuvre that the interlock table forbids is refused with the same indication a real
bay controller returns. These are declared design rules; a deployment would take them
from the operator's own interlock tables.

\subsection{Physical process validation}
\label{sec:gov-physical}

Every measurement the decoy exposes comes from a converged power flow on the
adversary's instance or from a transient run on it; the renderer never invents a
value. We call this physical process validation, and the per-adversary instance is
what makes it enforceable: the measurement an adversary reads is the solution of its
own manoeuvres, so it cannot contradict itself between two of its sessions. The one
case that needs a policy is non-convergence. When a manoeuvre leaves the instance
without a solution, the solver is re-run from a flat start and the topology is checked
for islands. A failure that survives both is not thereby shown to be numerical, since
a flat start can itself fail to converge \cite{pandapower2018}. The design records it
as indeterminate: the instance reverts to its last converged state, the command is
answered as a real gateway answers a rejected operation, and the event is logged for
the operator as unresolved. An island left without supply or a limit violated by a
converged solution is a physical event. For these the decoy shows what a real grid
would show, the undervoltage or overload trip that a declared protection model assigns
to the affected feeders, with the breaker-open indications and alarm entries the trip
produces; the instance is then re-solved with those feeders out of service.
Both cases are logged as invariant events rather than concealed. The protection model
and its validation by control-room operators belong to the agenda of
Section~\ref{sec:limitations}.

\subsection{Budgets and escalation}
\label{sec:gov-budgets}

Each adversary carries a budget in the units this article measures, watt-hours of
inference per session, watt-hours of simulation, megabytes of state and analyst
minutes, each a term of Equation~\ref{eq:der}. A session that exhausts its budget is
closed or downgraded to the deterministic core; a session whose continuity score or
command pattern crosses a declared threshold is escalated to a human. The numbers of
Section~\ref{sec:energy} set the scale: with thinking off a defender spends two to
three watt-hours to hold an agent through 25 commands, with thinking on 8 to 17 times
more, so the budget should be set so that the second configuration is never the
default.

\section{Limitations and research agenda}
\label{sec:limitations}

The limitations fall into three groups: the reach of the measurement, the
configuration choices it declares, and the status of the grid design. The attacker's
energy is a laboratory quantity. Both GPUs are ours, the boundary of the measurement is
the GPU, and the CPU-side energy of the microVM, the relay and the emulators is
counted on neither side, nor in the annual sum of Section~\ref{sec:worked}. The
subtraction of $P_{\mathrm{idle}}$ carries an uncertainty of about 1~W per sample
that the intervals do not propagate; the power traces of the campaign were not
retained, so it cannot be propagated after the fact, and only the traces of the 31
re-run sessions are released with the data. In the field the only claim we make is a
conditional estimate of floating-point work from the tokens observed and the model
class inferred. The measurement covers eight model pairs with five runs per
configuration, and a treatment arm on three pairs. Six cells straddle one and would
need more runs to decide. Five runs per cell support a conditional interval, not a
test of equivalence: no change with the injection level is detected, and the intervals
bound that change to a few percent per level without showing it to be zero.

Two configuration choices are declared rather than corrected. Thinking mode was
switched on both sides together, so the design cannot separate a defender that
reasons from an attacker that reasons, and with it on the defender's decisions ran up
to their token budget. The blind attacker of one pair hit its generation cap on nearly
every command, reproduced the behaviour under explicit stop sequences, and is excluded
from the pooled trends. A third choice is a definition. The diagnostic calls are
excluded from the attack cost by construction of the test bench and their energy is
removed in proportion to generated tokens. On 31 sessions re-run after the campaign
with per-call power traces, this allocation reproduces the measured split of the call
energy within about 2.5 percentage points, with no consistent direction. The raw traces of
the campaign itself were not retained, the validated sessions come from one pair with
a low diagnostic share, and the sign of the exchange survives the lower bound in 40 of
48 cells under the assumptions that define that bound. The calls also add wall-clock
time between commands, which the defender reads as a signal, and their effect on its
decisions was not isolated. Charging them to the attacker instead would raise the
joule ratio by up to 3.4 for Llama and 2.7 for Phi-4, so both accountings are
reported and the defined one is quoted. The replicate of
Section~\ref{sec:energy-results} exposed a dependence the campaign could not see.
Under the same declared settings, eight days apart, the Qwen3.5 defender's answer
after its reasoning fell from about 900 tokens to about 40 and its skipped decisions
disappeared. Its energy fell by a third, and the pair's thinking-on exchange moved
from 0.64--0.72 to 0.84--1.07, while a Gemma 4 session re-run on the same days
matched its campaign cell. The cause was not identified. The inference runtime
updates its llama.cpp engine pack on its own, the version in force at the start of the
campaign was not recorded, and the handling of the end of turn by that engine is the
suspected variable (\ref{app:energy}). Both measurements are reported, and the
thinking-on result for this pair is the range between them rather than either point:
the five-run interval of a cell measures its repeatability within a campaign, not
across the runtime's behaviour. Sessions are fixed at 25 commands, so the measurement
is cost at equal command count rather than at equal dwell; the per-address state of
the testbed is present but not exercised by single sessions. With thinking on, one
defender coalesced in the campaign five to six of its 27 decisions per session into
the ones it did take, which its ratio does not show, and whether the coalesced
commands needed an immediate answer was not measured. $P_{\mathrm{idle}}$ was
measured once per model and machine rather than before and after each run, and the
attacker's machine carries system processes that raise its idle draw by a few watts,
subtracted by construction.

The case for a generative decoy rests on three arguments of different standing
(Section~\ref{sec:intro}). Coverage is a premise taken from the generative honeypot
literature, not a measurement of this article, and the comparison between a decoy with
and without a model is companion work; the countermeasure and the cost exchange are
measured, on one bench and one adversary class. The grid honeynet is a design. Its
cyber level exists as the SSH bench and the workshop prototype, neither integrated
with a grid emulator; its physical level is specified to the point where its cost is
measured on one network size and one machine; no end-to-end deployment is claimed,
and the interference isolation it promises is conditional on an identity assignment
that is not calibrated. The mission value of Equation~\ref{eq:dld} rests on one
illustrative scenario with a declared load, a public value of lost load and a
prevented fraction carried as a range; the sufficiency check is therefore a break-even
statement under declared inputs, robust to single errors and not to compounded ones,
and not a measurement. The continuity estimator is inherited and uncalibrated, and on
the grid the manoeuvre log is both its strongest indicator and a spoofable one. The
steady-state model excludes fast protection and electromagnetic transients, and an
adversary who probes at those time scales is out of scope.

The agenda follows from the limitations. A design that switches thinking on one side
at a time would separate the defender's reasoning from the attacker's, with the
runtime's engine pack pinned and its version logged with every run. Runs on the pairs
with a high diagnostic share are logged with per-call traces to validate the
allocation where it weighs most, and the same runs, repeated without the diagnostic
calls or with the interval matched, isolate the effect of those calls on the
defender's decisions. Command-level timestamps and decision coverage are
reconstructed so that a missed action, not a missed decision, becomes the measured
quantity, and runs are added where the intervals straddle one. The autonomous attacker
is pointed at the Cowrie-based decoy, so that the two implemented systems are compared
on the same adversary; the stateless-against-stateful arm that the workshop paper
asked for is added in companion work. The physical level is deployed against the
emulators and the command path measured end to end, and assumed cooling and
simulation energy are replaced by measurements. The mission value of
Section~\ref{sec:mission} is extended from one illustrative scenario to a
probabilistic risk model on an operator's own network, with the telemetry, the
interlocks and the trip patterns validated by its control-room staff.

\section{Conclusion}
\label{sec:conclusion}

Cyber deception has been judged by realism and by detection for two decades. The
workshop paper this article extends argued that it should also be judged by the cost
it makes the adversary spend, and it measured one side of that exchange. This article
measures both, in the unit a critical-infrastructure operator pays for, for the
adversary that is now arriving: a language-model agent. Across eight model pairs and
330 runs, an autonomous language-model attacker spends more energy than the
language-model defender that holds it in 41 of the 48 configurations run with thinking
off. The token and joule asymmetries of the same engagement diverge, and the
divergence favours the defender in 42 of 45 cells: token counts hide the exchange.
Counter-injection, the one countermeasure that exists only against this
adversary, makes the attacker read more with no detectable change in the energy of
either side. With thinking on, switched on both sides at once, the defender pays 8 to
17 times more and the attacker 3 to 11, and for one pair the exchange is lost in the
campaign. A replicate of that pair reproduced the attacker's side and not the
defender's, and left the exchange below parity for the blind attacker and close to it
for the reasoning one. The thinking setting moves the defender's cost by an order of
magnitude and the sign of the result with it, and which setting is in force, on both
sides, is part of what a measurement of this kind must record.

For the grid the article adds a design rather than a deployment. In the two-level
stateful architecture each identified adversary owns a private steady-state power
flow, so that concurrent attackers never see each other's manoeuvres, at a cost of the
same order as the digital state a stateful decoy already keeps. The model decides and
renders; the physical process and the protocols stay out of its hands. The exchange
rate is restated with a denominator whose inference term is measured, and completed
with the mission value the deception must protect, reduced to a break-even product. A
decoy that costs a few thousand euro a year passes the test whenever the likelihood of
a single three-hour substation outage, multiplied by the fraction of that loss it
prevents, exceeds about one percent. Whether the defence is affordable is that
calculation, not the energy ratio alone. The adversary's energy is a laboratory
measurement; in the field the defender can only estimate the adversary's
floating-point work, conditionally, not its energy. The honeynet is designed, not
built, and the comparison between a stateful and a stateless decoy belongs to
companion work. In this bench, when both sides are language models and the defender
keeps its answers short, the attacker's marginal GPU energy exceeds the defender's
under the declared accounting, and the operator can say by how much.

\section*{CRediT authorship contribution statement}
\textbf{Vlad Mihal:} Software, Investigation, Data curation.
\textbf{Hosam Alamleh:} Conceptualization, Formal analysis, Supervision, Writing -- review \& editing.
\textbf{Alessandro Cantelli-Forti:} Conceptualization, Methodology, Formal analysis,
Writing -- original draft, Writing -- review \& editing, Supervision.

\section*{Declaration of competing interest}
Alessandro Cantelli-Forti is an author of the 2018
stateful-honeypot architecture reused here \cite{cantelliforti2018}. The authors
declare no other competing financial interests or personal relationships that could
have influenced the work reported in this article.

\section*{Data availability}
The per-run energy, token and floating-point data of the 330 sessions of
Section~\ref{sec:energy} and of the 30-session replicate, the power traces of the 31
re-run sessions used for the validation of Section~\ref{sec:energy-honesty} (the
replicate among them; the traces of the campaign itself were not retained,
Section~\ref{sec:limitations}), the scripts that generate every table and figure of
that section, the validation script and the benchmark of \ref{app:pandapower} with its
raw results are deposited at Zenodo~\cite{masm2026data},
\url{https://doi.org/10.5281/zenodo.23101898}, under a CC BY 4.0 licence. The command
text of the sessions is withheld; the per-call files carry its length and digest. The
supplementary material at the end of this preprint tabulates every cell
under both accountings, the ultra-over-off ratios, the treatment arm and the
per-session validation, and adds the per-cell heat maps and the resource profile of
every pair.

\section*{Declaration of generative AI and AI-assisted technologies in the writing process}
During the preparation of this work the authors used Claude (Anthropic) and Codex
(OpenAI) to proofread the text and to write the scripts that generate the tables and
figures from the per-run data. After using these tools the authors re-edited the
content and take full responsibility for the content of the publication.

\section*{Acknowledgments}
The authors thank Carlo Alberto Nucci and Eleonora Riva Sanseverino for discussions on
service-interruption scenarios and on the value of lost load, which shaped
Sections~\ref{sec:framework-mission} and~\ref{sec:mission}.

\appendix
\renewcommand{\thesection}{Appendix~\Alph{section}}
\renewcommand{\thesubsection}{\Alph{section}.\arabic{subsection}}
\renewcommand{\thetable}{\Alph{section}.\arabic{table}}
\renewcommand{\thefigure}{\Alph{section}.\arabic{figure}}
\section{Energy measurement protocol}
\label{app:energy}

The protocol has three parts: measurement, attribution and settings. Energy on each
side is the integral of GPU power net of idle over the session,
$E = \sum_t (P_t - P_{\mathrm{idle}})\,\Delta t$, from the nvidia-smi log sampled once per
second (power draw, memory used, compute and memory-controller utilisation).
$P_{\mathrm{idle}}$ is measured once per model and per machine before its sessions, with
the model loaded and the same system processes running, over windows of 25 to 96~s in
which the idle draw stayed within 1~W; the integrand is clipped at zero per sample, and
an error of 1~W on $P_{\mathrm{idle}}$ is the uncertainty of the subtraction. A
calibration run of continuous generation before each pair gives 1.6 to 5.7~J per
generated token, against 2 to 13~J per generated token within sessions, which include
prompt reading and the gaps between calls. The calibration is a plausibility check, and
every result rests on the session energy.

Attribution follows the declared accounting. Tokens are prompt plus completion as
reported by the inference server and are attributed per call. Floating-point work is
$2NT$ with $N$ the active parameters, so that the mixture-of-experts models are charged
for the experts they use rather than for their total size. The attack cost is the cost
of the command calls alone (Section~\ref{sec:energy-honesty}). The diagnostic calls the
harness makes after each command are excluded from tokens, floating-point work and
energy, and their energy is removed in proportion to generated tokens. Energy per token
divides the energy of the command calls by their completion tokens on the attacker
side, and the session energy by the completion tokens on the defender side. Attacker
and defender are variants of the same model with the same active parameters, so the
floating-point ratio equals the token ratio and is not reported separately. Ratios are
computed per run and summarised by the geometric mean with a 95\% interval on the log
scale; the per-configuration summaries of the reports use the ratio of means, a
different estimand that exceeds the geometric mean by up to 20 percent in the most
dispersed cell.

The inference settings below apply to both sides, with caps and temperatures set per
kind of call. Both sides run LM Studio 0.4.24 with a context length of 50,000 tokens.
Generation caps are 3,000 tokens for the attacker's command call and for the defender's
decision, and 220 for the diagnostic call; the temperature is 0.4, 0.1 and 0.0 for the
three kinds of call, set by the client, and the other sampling parameters stay at the
preset defaults of each model. The main set runs with the runtime's thinking mode off.
The treatment arm runs the three pairs that support it with thinking on, under a
reasoning budget of 1,024 tokens and the native stop sequences of each family (Qwen:
\texttt{<|im\_end|>}, \texttt{<|endoftext|>}; Gemma: \texttt{<end\_of\_turn>},
\texttt{<eos>}; GLM its own).

The two hosts are HP Z2 Tower G9 workstations with an Intel Core i9-13900K, 128~GB of
DDR5 memory, a 2~TB NVMe drive and one NVIDIA RTX 5000 Ada GPU with 32~GB, under
Windows 11 Pro, the same machines as in the workshop paper. Inference runs on the GPU
and the energy is read from the GPU alone, so the host's other components do not enter
the measured quantities. Both hosts serve their model through the local
OpenAI-compatible endpoint of LM Studio 0.4.24. The runtime's automatic loading
settings, which set the number of layers offloaded to the GPU, the batch size and the
cache handling from the model file and the available GPU memory, were left at their
defaults and are the same on both hosts; the sampling parameters the client does not
set come from the metadata of each model file \cite{mihal2026}. LM Studio loads the
GGUF models through a separately versioned llama.cpp engine pack, which it updates on
its own when automatic updates are enabled, as they were on both hosts. At the end of
the campaign both hosts ran the CUDA~12 llama.cpp engine 2.44.0 on NVIDIA driver 596.86
with CUDA~13.2; the version in force at the start of the campaign was not recorded.
Table~\ref{tab:models} names the model files of the eight pairs as the reports name
them; the SHA-256 digests of the model files are released with the data.

\begin{table}[h]
\centering
\caption{Model files of the eight pairs, as the run reports name them.}
\label{tab:models}
\scriptsize\setlength{\tabcolsep}{3pt}
\begin{tabular}{@{}ll@{}}
\toprule
Defender (stock) & Attacker (uncensored or abliterated) \\
\midrule
\path{qwen2.5-coder-7b-instruct} & \path{qwen2.5-coder-7b-instruct-uncensored} \\
\path{bartowski/meta-llama-3.1-8b-instruct@q4_k_m} & \path{llama-3.1-8b-lexi-uncensored-v2@q4_k_m} \\
\path{qwen/qwen3.5-9b} & \path{qwen3.5-9b-uncensored-hauhaucs-aggressive} \\
\path{mistralai/ministral-3-14b-reasoning} & \path{ministral-3-14b-reasoning-2512-uncensored-i1} \\
\path{phi-4} & \path{phi-4-uncensored} \\
\path{google/gemma-4-26b-a4b} & \path{gemma4-26b-a4b-uncensored-hauhaucs-balanced} \\
\path{zai-org/glm-4.7-flash} & \path{glm-4.7-flash-uncensored-hauhaucs-aggressive} \\
\path{qwen/qwen3-30b-a3b-2507} & \path{huihui-qwen3-30b-a3b-instruct-2507-abliterated} \\
\bottomrule
\end{tabular}
\end{table}

The per-cell results under both accountings, the ultra-over-off ratios and the
treatment arm are tabulated in the supplementary material (Tables~S1 to S3), together
with the joule and token ratios per cell as heat maps and the GPU utilisation, peak
memory and calibration profile of every pair (Figures~S1 and S2); all of them are
generated from the per-run data by the released scripts.


\section{Per-adversary power flows: cost measurement}
\label{app:pandapower}

The measurement answers one question of Section~\ref{sec:arch-physical}: what it
costs, in latency and memory, to keep $N$ private steady-state power flows resident,
one per identified adversary, and to update one of them when its owner issues a
breaker or line command. It covers $N \in \{1, 10, 100, 1000\}$ under the two
representations of $G_i$ of Section~\ref{sec:arch-state}: a full copy, one network
object per adversary, and a delta, one shared scratch network plus, per adversary, the
set of elements it has switched and a cached copy of its last result tables. The base
case is the IEEE 39-bus New England system as shipped with \pandapower{} 3.5.4, 39
buses, 35 lines and 11 transformers. A manoeuvre toggles the in-service state of one
line chosen at random with a fixed seed and re-solves the instance with the
Newton--Raphson solver, numba enabled; a manoeuvre whose power flow does not converge
is counted, reverted and left out of the latency statistics, and the delta
representation measured here keeps the set of elements each adversary has switched,
not the ordered log of Section~\ref{sec:arch-state}. Fork is the creation of $G_i$ for
a new adversary, timed once per instance. Update is one manoeuvre and one power flow
on a resident instance chosen at random, 50 attempts after 5 warm-ups, of which 45,
47, 50 and 50 converged at $N = 1$, 10, 100 and 1000 under both representations. The
concurrent update runs two worker threads that each apply 25 manoeuvres in a row to a
different instance, 41 to 43 of the 50 converging because the outages accumulate on
the same instance; the delta's solves serialise on the lock that guards its shared
scratch network. Memory per instance is the Python allocation traced after the base
case and the scratch network exist, numpy included, divided by $N$, once every
instance carries a manoeuvre and a cached solution: the increment per instance, not
the memory of the process. Latencies are medians and 95th percentiles over the
converged attempts; memory is in MiB. The machine is a Linux container with two
virtual cores of an Intel Xeon processor at 2.80~GHz and 7.8~GB of memory, Python
3.11.15, numpy 2.4.4; the base case solves in 35~ms on it. The script, the raw
results and the environment record are released with the data.

\begin{table}[h]
\centering
\scriptsize\setlength{\tabcolsep}{2.5pt}
\caption{Per-adversary power-flow cost on the IEEE 39-bus case with \pandapower{}
3.5.4 on two virtual cores. Medians and 95th percentiles over the converged attempts
of 50; throughput in converged updates per second with two workers; memory per
resident instance once every instance carries a manoeuvre and a cached solution (at
$N = 1$ the delta figure includes the allocations of the shared scratch network's
first solve).}
\label{tab:cost}
\IfFileExists{fig/cost_table.tex}{
\begin{tabular}{@{}lrrrrrr@{}}
\toprule
$G_i$ & $N$ & Fork & Update & Update, 2 workers & Throughput & Memory \\
 & & (ms) & median (p95), ms & median (p95), ms & (updates/s) & (MiB/inst.) \\
\midrule
full copy & 1 & 12.3 & 35.4 (48.5) & -- & -- & 0.57 \\
full copy & 10 & 11.0 & 33.6 (42.9) & 60.8 (177.0) & 17 & 0.59 \\
full copy & 100 & 11.5 & 36.2 (49.4) & 83.8 (120.8) & 19 & 0.58 \\
full copy & 1000 & 11.7 & 35.5 (50.3) & 72.4 (126.2) & 20 & 0.58 \\
\midrule
delta & 1 & 0.0 & 39.0 (51.3) & -- & -- & 0.26 \\
delta & 10 & 0.0 & 34.4 (36.9) & 35.7 (45.3) & 22 & 0.04 \\
delta & 100 & 0.0 & 34.1 (41.6) & 38.2 (56.0) & 20 & 0.03 \\
delta & 1000 & 0.0 & 33.7 (39.5) & 34.3 (47.0) & 24 & 0.02 \\
\bottomrule
\end{tabular}
}{%
  \textcolor{red}{fig/cost\_table.tex missing: run data/make\_cost\_table.py}}
\end{table}

Table~\ref{tab:cost} gives the result. An update costs 34 to 39~ms at the median and
37 to 51~ms at the 95th percentile, the same from one to a thousand resident
instances under both representations, because the solver, not the bookkeeping, sets the
time. A full copy costs 11 to 12~ms to fork and 0.58~MiB to keep; a delta forks in
microseconds and keeps 0.02 to 0.04~MiB. With two workers updating different instances
at once the delta keeps its latency, 34 to 38~ms at the median and 45 to 56 at the
95th percentile, while the full copy's rises to 61 to 84~ms (121 to 177), at 17 to 21
converged updates per second for the full copy and 20 to 24 for the delta: on two
cores the full copies contend for shared compute resources, while the delta
serialises its solves on the shared scratch network. A thousand resident adversaries
are therefore
22~MiB of state as deltas or 0.6~GiB as full copies. Whether the solver stays mostly
idle depends on the aggregate command rate and on the policy for background updates:
at the few commands per minute per session of Section~\ref{sec:energy}, two cores
serve a few hundred concurrently active sessions, and a thousand resident states need
not be active at once. The measurement covers one network size, one machine and one
update per manoeuvre; the latency scales with the network, tens of milliseconds for a
few hundred buses in \pandapower{} \cite{pandapower2018}, and the memory per instance
with the size of its result tables.

\renewcommand{\thesection}{\arabic{section}}
\bibliographystyle{elsarticle-num}
\bibliography{refs}

\clearpage
\setcounter{table}{0}
\setcounter{figure}{0}
\renewcommand{\thetable}{S\arabic{table}}
\renewcommand{\thefigure}{S\arabic{figure}}
\section*{Supplementary material}

\noindent
This supplementary material tabulates the per-cell results of the energy measurement of
Section~\ref{sec:energy}, the ultra-over-off ratios, the treatment arm and the
per-session validation of the token allocation, and it adds two figures: the joule and
token ratios per cell as heat maps, and the GPU utilisation, peak memory and calibration
profile of every pair. Every table and figure is generated from the per-run data by the
released scripts (\path{make_energy_tables.py}, \path{make_energy_figures2.py},
\path{make_validation_table.py}); nothing is typed by hand. Notation, flags and
accountings are those of the article: $\rhoJ$ is the attacker-to-defender ratio of
marginal GPU energy per session, net of idle; the defined accounting removes the energy
of the attacker's diagnostic calls in proportion to generated tokens and the total
accounting keeps every joule the attacking GPU drew; $\rhotok$ is the ratio of prompt
plus completion tokens; flag X marks the truncated blind attacker of Qwen3 30B-A3B and
flag T the runs with thinking on. Ratios are summarised by the geometric mean with a
95\% interval on the log scale ($t$ distribution, $n-1$ degrees of freedom).

\begin{sidewaystable}[p]
\centering
\scriptsize\setlength{\tabcolsep}{3pt}
\caption{Energy and token exchange per cell of the main set (model pair, attacker type,
injection level, thinking off), five runs each. $E$ per session in Wh net of idle; diag.\
share is the fraction of attacker energy the reports attribute to diagnostic calls;
$\rhoJ$ under the total and the pure (defined) accounting, geometric mean with 95\%
interval; $\rhotok$ geometric mean.}
\label{tab:s1}
\IfFileExists{fig/energy_cells.tex}{
\begin{tabular}{@{}llllrrrrllrl@{}}
\toprule
Model & Att. & Inj. & $n$ & $\ED$ & $\EA^{\mathrm{tot}}$ & $\EA^{\mathrm{pure}}$ & diag. & $\rhoJ^{\mathrm{tot}}$ & $\rhoJ^{\mathrm{pure}}$ & $\rhotok$ & Flag \\
 & & & & (Wh) & (Wh) & (Wh) & share & [CI95] & [CI95] & & \\
\midrule
Qwen2.5-Coder 7B & blind & off & 5 & 2.1 & 11.6 & 10.8 & 0.09 & 4.66 [2.28--9.50] & 4.22 [1.87--9.51] & 1.80 &  \\
Qwen2.5-Coder 7B & blind & progr. & 5 & 1.9 & 5.6 & 4.6 & 0.22 & 2.80 [1.88--4.17] & 2.17 [1.21--3.89] & 1.30 &  \\
Qwen2.5-Coder 7B & blind & ultra & 5 & 1.9 & 4.9 & 4.0 & 0.19 & 2.58 [2.06--3.22] & 2.08 [1.55--2.80] & 1.54 &  \\
Qwen2.5-Coder 7B & reasoning & off & 5 & 1.7 & 2.5 & 1.6 & 0.38 & 1.54 [1.44--1.65] & 0.95 [0.91--1.00] & 0.96 &  \\
Qwen2.5-Coder 7B & reasoning & progr. & 5 & 1.6 & 2.8 & 1.9 & 0.33 & 1.73 [1.56--1.93] & 1.16 [0.97--1.39] & 1.28 &  \\
Qwen2.5-Coder 7B & reasoning & ultra & 5 & 1.6 & 3.0 & 2.0 & 0.34 & 1.92 [1.68--2.21] & 1.28 [1.05--1.55] & 1.46 &  \\
Llama 3.1 8B & blind & off & 5 & 2.1 & 6.8 & 2.2 & 0.67 & 3.26 [3.03--3.51] & 1.07 [0.91--1.26] & 0.72 &  \\
Llama 3.1 8B & blind & progr. & 5 & 2.1 & 8.5 & 2.5 & 0.71 & 4.04 [3.27--5.00] & 1.17 [0.92--1.50] & 1.08 &  \\
Llama 3.1 8B & blind & ultra & 5 & 2.0 & 8.7 & 2.7 & 0.69 & 4.37 [3.69--5.18] & 1.30 [0.91--1.87] & 1.15 &  \\
Llama 3.1 8B & reasoning & off & 5 & 1.7 & 6.0 & 2.6 & 0.56 & 3.54 [3.17--3.95] & 1.55 [1.36--1.75] & 0.88 &  \\
Llama 3.1 8B & reasoning & progr. & 5 & 1.9 & 8.2 & 2.7 & 0.68 & 4.45 [3.95--5.02] & 1.39 [0.95--2.04] & 1.27 &  \\
Llama 3.1 8B & reasoning & ultra & 5 & 1.9 & 8.2 & 3.2 & 0.62 & 4.32 [3.45--5.39] & 1.60 [1.00--2.57] & 1.17 &  \\
Qwen3.5 9B & blind & off & 5 & 2.4 & 7.3 & 5.5 & 0.25 & 3.01 [2.87--3.16] & 2.27 [2.04--2.52] & 1.47 &  \\
Qwen3.5 9B & blind & progr. & 5 & 2.4 & 9.9 & 7.7 & 0.25 & 3.82 [2.51--5.83] & 2.83 [1.62--4.96] & 1.75 &  \\
Qwen3.5 9B & blind & ultra & 5 & 2.6 & 9.9 & 7.8 & 0.22 & 3.76 [3.22--4.39] & 2.93 [2.38--3.61] & 1.75 &  \\
Qwen3.5 9B & reasoning & off & 5 & 2.2 & 10.8 & 9.7 & 0.11 & 5.06 [4.72--5.43] & 4.51 [4.17--4.88] & 1.05 &  \\
Qwen3.5 9B & reasoning & progr. & 5 & 2.2 & 11.6 & 10.4 & 0.11 & 5.20 [4.67--5.80] & 4.65 [4.09--5.30] & 1.17 &  \\
Qwen3.5 9B & reasoning & ultra & 5 & 2.4 & 12.0 & 10.7 & 0.11 & 5.06 [4.64--5.51] & 4.52 [4.11--4.97] & 1.26 &  \\
Ministral 3 14B R. & blind & off & 5 & 2.5 & 7.6 & 5.2 & 0.32 & 3.08 [2.72--3.50] & 2.10 [1.79--2.46] & 1.27 &  \\
Ministral 3 14B R. & blind & progr. & 5 & 2.5 & 7.9 & 5.4 & 0.32 & 3.04 [2.41--3.84] & 2.07 [1.54--2.80] & 1.28 &  \\
Ministral 3 14B R. & blind & ultra & 5 & 2.5 & 8.4 & 5.9 & 0.31 & 3.37 [3.07--3.71] & 2.32 [1.89--2.85] & 1.45 &  \\
Ministral 3 14B R. & reasoning & off & 5 & 2.5 & 6.8 & 4.9 & 0.27 & 2.76 [2.40--3.17] & 2.02 [1.70--2.38] & 1.04 &  \\
Ministral 3 14B R. & reasoning & progr. & 5 & 2.6 & 8.0 & 5.8 & 0.28 & 3.09 [2.59--3.68] & 2.22 [1.85--2.67] & 1.44 &  \\
Ministral 3 14B R. & reasoning & ultra & 5 & 2.4 & 7.2 & 5.3 & 0.27 & 3.01 [2.61--3.49] & 2.21 [1.92--2.54] & 1.18 &  \\
Phi-4 15B & blind & off & 5 & 2.7 & 10.8 & 4.1 & 0.62 & 4.05 [3.25--5.03] & 1.50 [1.06--2.12] & 0.79 &  \\
Phi-4 15B & blind & progr. & 5 & 2.7 & 13.1 & 6.9 & 0.46 & 4.84 [4.23--5.55] & 2.54 [2.05--3.15] & 1.23 &  \\
Phi-4 15B & blind & ultra & 5 & 2.7 & 11.8 & 4.8 & 0.58 & 4.36 [3.49--5.45] & 1.81 [1.62--2.01] & 1.21 &  \\
Phi-4 15B & reasoning & off & 5 & 2.8 & 20.5 & 12.2 & 0.41 & 7.19 [6.79--7.61] & 4.26 [3.74--4.85] & 0.95 &  \\
Phi-4 15B & reasoning & progr. & 5 & 2.7 & 18.0 & 11.7 & 0.35 & 6.56 [5.38--7.99] & 4.27 [3.72--4.91] & 1.08 &  \\
Phi-4 15B & reasoning & ultra & 5 & 2.9 & 17.9 & 12.3 & 0.31 & 6.15 [5.51--6.86] & 4.21 [3.44--5.16] & 1.22 &  \\
Gemma 4 26B-A4B & blind & off & 5 & 2.1 & 10.9 & 5.6 & 0.50 & 5.09 [4.34--5.97] & 2.50 [1.63--3.84] & 1.68 &  \\
Gemma 4 26B-A4B & blind & progr. & 5 & 2.2 & 12.7 & 7.3 & 0.43 & 5.81 [4.70--7.19] & 3.29 [2.31--4.67] & 1.97 &  \\
Gemma 4 26B-A4B & blind & ultra & 5 & 2.1 & 12.5 & 7.1 & 0.43 & 6.07 [5.58--6.60] & 3.44 [2.87--4.12] & 2.09 &  \\
Gemma 4 26B-A4B & reasoning & off & 5 & 1.9 & 11.2 & 7.3 & 0.35 & 5.82 [5.29--6.40] & 3.78 [3.23--4.41] & 1.12 &  \\
Gemma 4 26B-A4B & reasoning & progr. & 5 & 1.8 & 11.6 & 7.7 & 0.34 & 6.57 [6.08--7.10] & 4.34 [3.94--4.78] & 1.32 &  \\
Gemma 4 26B-A4B & reasoning & ultra & 5 & 1.8 & 11.6 & 7.6 & 0.34 & 6.54 [6.07--7.04] & 4.30 [3.93--4.69] & 1.43 &  \\
GLM-4.7 Flash 30B-A3B & blind & off & 5 & 2.0 & 11.1 & 8.1 & 0.28 & 5.56 [4.46--6.94] & 4.00 [2.78--5.76] & 0.88 &  \\
GLM-4.7 Flash 30B-A3B & blind & progr. & 5 & 2.1 & 9.8 & 6.7 & 0.32 & 4.76 [4.23--5.36] & 3.23 [2.67--3.91] & 0.79 &  \\
GLM-4.7 Flash 30B-A3B & blind & ultra & 5 & 2.2 & 7.7 & 4.4 & 0.44 & 3.57 [2.94--4.33] & 1.97 [1.38--2.80] & 0.76 &  \\
GLM-4.7 Flash 30B-A3B & reasoning & off & 5 & 2.0 & 7.7 & 4.6 & 0.42 & 3.85 [3.45--4.30] & 2.24 [1.75--2.86] & 0.89 &  \\
GLM-4.7 Flash 30B-A3B & reasoning & progr. & 5 & 2.1 & 6.5 & 3.0 & 0.54 & 3.18 [2.91--3.47] & 1.46 [1.19--1.78] & 0.92 &  \\
GLM-4.7 Flash 30B-A3B & reasoning & ultra & 5 & 2.0 & 6.7 & 3.8 & 0.44 & 3.28 [2.81--3.83] & 1.83 [1.44--2.31] & 0.84 &  \\
Qwen3 30B-A3B & blind & off & 5 & 1.9 & 82.0 & 76.4 & 0.07 & 44.07 [40.98--47.40] & 41.08 [38.23--44.14] & 8.92 & X \\
Qwen3 30B-A3B & blind & progr. & 5 & 1.8 & 83.9 & 78.1 & 0.07 & 46.17 [43.57--48.92] & 43.01 [40.59--45.58] & 9.16 & X \\
Qwen3 30B-A3B & blind & ultra & 5 & 1.9 & 82.1 & 76.7 & 0.07 & 44.72 [42.77--46.76] & 41.76 [40.07--43.53] & 8.93 & X \\
Qwen3 30B-A3B & reasoning & off & 5 & 1.7 & 16.9 & 14.1 & 0.17 & 10.05 [9.05--11.16] & 8.39 [7.39--9.52] & 1.46 &  \\
Qwen3 30B-A3B & reasoning & progr. & 5 & 1.6 & 14.9 & 12.1 & 0.19 & 9.06 [8.41--9.77] & 7.35 [6.71--8.06] & 1.26 &  \\
Qwen3 30B-A3B & reasoning & ultra & 5 & 1.7 & 14.8 & 12.1 & 0.19 & 8.90 [8.02--9.87] & 7.24 [6.53--8.03] & 1.31 &  \\
\bottomrule
\end{tabular}
}{\textcolor{red}{fig/energy\_cells.tex missing}}
\end{sidewaystable}

\begin{sidewaystable}[p]
\centering
\footnotesize\setlength{\tabcolsep}{2.4pt}
\caption{What deception moves: ultra over off, per model pair and attacker type, main
set with thinking off. Geometric-mean ratios with Welch 95\% intervals on the log scale;
five runs per level. Attacker energy under the defined accounting. X marks the truncated
blind attacker.}
\label{tab:s2}
\IfFileExists{fig/energy_rr.tex}{
\begin{tabular}{@{}llllllll@{}}
\toprule
Model & Att. & $\rhoJ$ & $\rhotok$ & Att.\ tokens & Att.\ energy & Def.\ energy & Flag \\
 & & [CI95] & [CI95] & [CI95] & [CI95] & [CI95] & \\
\midrule
Qwen2.5-Coder 7B & blind & 0.49 [0.22--1.10] & 0.85 [0.46--1.57] & 0.78 [0.39--1.56] & 0.43 [0.17--1.07] & 0.87 [0.76--0.99] &  \\
Qwen2.5-Coder 7B & reasoning & 1.34 [1.11--1.62] & 1.53 [1.30--1.79] & 1.51 [1.29--1.77] & 1.26 [1.01--1.57] & 0.94 [0.85--1.03] &  \\
Llama 3.1 8B & blind & 1.21 [0.85--1.73] & 1.60 [1.29--1.99] & 1.59 [1.29--1.94] & 1.16 [0.80--1.68] & 0.96 [0.92--1.00] &  \\
Llama 3.1 8B & reasoning & 1.04 [0.65--1.65] & 1.34 [1.01--1.78] & 1.34 [1.01--1.76] & 1.15 [0.72--1.82] & 1.11 [0.93--1.32] &  \\
Qwen3.5 9B & blind & 1.29 [1.05--1.59] & 1.19 [0.90--1.59] & 1.22 [0.94--1.59] & 1.40 [1.13--1.74] & 1.08 [1.02--1.15] &  \\
Qwen3.5 9B & reasoning & 1.00 [0.90--1.11] & 1.20 [1.02--1.42] & 1.22 [1.06--1.40] & 1.11 [1.01--1.22] & 1.11 [1.05--1.17] &  \\
Ministral 3 14B R. & blind & 1.11 [0.89--1.38] & 1.14 [1.03--1.26] & 1.13 [1.02--1.25] & 1.12 [0.82--1.53] & 1.01 [0.89--1.14] &  \\
Ministral 3 14B R. & reasoning & 1.09 [0.91--1.31] & 1.14 [0.88--1.48] & 1.16 [0.90--1.49] & 1.07 [0.86--1.33] & 0.98 [0.90--1.06] &  \\
Phi-4 15B & blind & 1.20 [0.86--1.69] & 1.53 [1.23--1.89] & 1.53 [1.20--1.94] & 1.20 [0.87--1.67] & 1.00 [0.94--1.06] &  \\
Phi-4 15B & reasoning & 0.99 [0.80--1.22] & 1.28 [1.03--1.60] & 1.30 [1.05--1.61] & 1.01 [0.82--1.25] & 1.02 [0.90--1.16] &  \\
Gemma 4 26B-A4B & blind & 1.37 [0.90--2.09] & 1.25 [1.05--1.48] & 1.29 [1.06--1.58] & 1.32 [0.85--2.05] & 0.96 [0.93--0.99] &  \\
Gemma 4 26B-A4B & reasoning & 1.14 [0.97--1.33] & 1.28 [0.99--1.64] & 1.24 [0.93--1.65] & 1.04 [0.87--1.25] & 0.92 [0.87--0.97] &  \\
GLM-4.7 Flash 30B-A3B & blind & 0.49 [0.32--0.75] & 0.86 [0.68--1.08] & 0.92 [0.70--1.20] & 0.54 [0.36--0.80] & 1.09 [0.97--1.22] &  \\
GLM-4.7 Flash 30B-A3B & reasoning & 0.82 [0.61--1.09] & 0.95 [0.81--1.11] & 0.94 [0.80--1.12] & 0.83 [0.62--1.11] & 1.01 [0.96--1.06] &  \\
Qwen3 30B-A3B & blind & 1.02 [0.95--1.09] & 1.00 [0.96--1.04] & 0.99 [0.96--1.03] & 1.00 [0.95--1.06] & 0.99 [0.94--1.03] & X \\
Qwen3 30B-A3B & reasoning & 0.86 [0.75--0.99] & 0.90 [0.75--1.08] & 0.90 [0.75--1.07] & 0.86 [0.73--1.00] & 0.99 [0.95--1.03] &  \\
\bottomrule
\end{tabular}
}{\textcolor{red}{fig/energy\_rr.tex missing}}
\end{sidewaystable}

\begin{sidewaystable}[p]
\centering
\footnotesize\setlength{\tabcolsep}{2.4pt}
\caption{Treatment arm: the three pairs whose models support a thinking mode, run with
it off (main set) and on, 15 runs per row (three injection levels), and the replicate
of the Qwen3.5 thinking-on cells of 1 October 2026 (rows ``on, replicate''). Defender completion
tokens per decision, decisions per session, attacker completion tokens per command, $E$
the marginal GPU energy per session in Wh net of idle (attacker: defined allocation),
means; $\rhoJ$ under the defined accounting, geometric mean with 95\% interval;
$\rhotok$ geometric mean.}
\label{tab:s3}
\IfFileExists{fig/energy_thinking.tex}{
\begin{tabular}{@{}lllrrrrrrlr@{}}
\toprule
Model & Att. & Think. & $n$ & Def.\ tok./dec. & Dec./sess. & Att.\ tok./cmd & $\ED$ & $\EA^{\mathrm{pure}}$ & $\rhoJ^{\mathrm{pure}}$ & $\rhotok$ \\
 & & & & & & & (Wh) & (Wh) & [CI95] & \\
\midrule
Qwen3.5 9B & blind & off & 15 & 50 & 26.9 & 181 & 2.5 & 7.0 & 2.66 [2.27--3.12] & 1.65 \\
Qwen3.5 9B & blind & on & 15 & 1,889 & 20.6 & 871 & 33.1 & 22.2 & 0.67 [0.62--0.73] & 1.19 \\
Qwen3.5 9B & blind & on, replicate & 15 & 1,068 & 26.1 & 886 & 26.0 & 22.3 & 0.86 [0.82--0.89] & 1.01 \\
Qwen3.5 9B & reasoning & off & 15 & 48 & 26.9 & 387 & 2.3 & 10.3 & 4.56 [4.37--4.76] & 1.15 \\
Qwen3.5 9B & reasoning & on & 15 & 1,993 & 21.6 & 1,128 & 38.6 & 27.0 & 0.70 [0.67--0.73] & 1.06 \\
Qwen3.5 9B & reasoning & on, replicate & 15 & 1,062 & 26.0 & 1,152 & 25.9 & 27.3 & 1.05 [1.02--1.08] & 1.10 \\
Gemma 4 26B-A4B & blind & off & 15 & 32 & 27.0 & 269 & 2.1 & 6.7 & 3.05 [2.60--3.58] & 1.91 \\
Gemma 4 26B-A4B & blind & on & 15 & 1,032 & 27.3 & 1,458 & 17.2 & 27.7 & 1.59 [1.45--1.74] & 1.41 \\
Gemma 4 26B-A4B & reasoning & off & 15 & 33 & 27.0 & 420 & 1.8 & 7.5 & 4.13 [3.89--4.39] & 1.29 \\
Gemma 4 26B-A4B & reasoning & on & 15 & 1,036 & 26.0 & 1,405 & 16.8 & 24.1 & 1.43 [1.37--1.50] & 1.31 \\
GLM-4.7 Flash 30B-A3B & blind & off & 15 & 45 & 26.9 & 421 & 2.1 & 6.4 & 2.94 [2.38--3.64] & 0.81 \\
GLM-4.7 Flash 30B-A3B & blind & on & 15 & 1,909 & 25.7 & 2,198 & 25.8 & 38.0 & 1.48 [1.32--1.66] & 1.03 \\
GLM-4.7 Flash 30B-A3B & reasoning & off & 15 & 53 & 27.0 & 226 & 2.0 & 3.8 & 1.81 [1.58--2.08] & 0.88 \\
GLM-4.7 Flash 30B-A3B & reasoning & on & 15 & 1,874 & 25.7 & 2,273 & 25.3 & 41.7 & 1.65 [1.49--1.82] & 1.05 \\
\bottomrule
\end{tabular}
}{\textcolor{red}{fig/energy\_thinking.tex missing}}
\end{sidewaystable}

\begin{table}[p]
\centering
\footnotesize\setlength{\tabcolsep}{4pt}
\caption{Validation of the token allocation against per-call power traces on the 30
sessions of the Qwen3.5 9B pair re-run with thinking on (1 October 2026, the replicate)
and on the single session of the same pair traced on 30 September (last row). For each
session the attacker's GPU power, sampled at 1~Hz and taken net of the pre-session
idle, is partitioned by the harness timestamps, every sample counted once: a diagnostic
call runs from the instant the command's output is received to the diagnostic's own
timestamp; the next command's call runs from that instant until the GPU power first
falls below 30~W net after half of its reported generation time, so that it includes
prompt processing and generation; what remains is the wait for the decoy's answer.
Diag.\ share and Wait share are the fractions of the session energy in the diagnostic
calls and in the waits; Token share is $T_{\mathrm{cmd}}/(T_{\mathrm{cmd}}+T_{\mathrm{diag}})$
on completion tokens, the share the allocation uses; Energy share is the measured
$E_{\mathrm{cmd}}/(E_{\mathrm{cmd}}+E_{\mathrm{diag}})$ of the call energy; Diff.\ is
their difference in percentage points, positive when the allocation overstates the
command energy. Sensitivity on the 30 sessions: with the threshold at 20 or 50~W the
differences change by less than 0.3 points; with the waits charged to the command call
they run from $-2.7$ to 2.1 (mean $-0.3$).}
\label{tab:s4}
\resizebox{\textwidth}{!}{\IfFileExists{fig/validation_table.tex}{
\begin{tabular}{@{}llrrrrrrr@{}}
\toprule
Attacker & Inj. & Run & $E_{\mathrm{trace}}$ (Wh) & Diag.\ share & Wait share & Token share & Energy share & Diff.\ (pp) \\
\midrule
blind & off & 1 & 29.6 & 0.176 & 0.030 & 0.816 & 0.819 & $-$0.3 \\
blind & off & 2 & 30.4 & 0.169 & 0.016 & 0.808 & 0.829 & $-$2.0 \\
blind & off & 3 & 26.7 & 0.210 & 0.004 & 0.801 & 0.789 & 1.2 \\
blind & off & 4 & 25.9 & 0.213 & 0.018 & 0.789 & 0.783 & 0.6 \\
blind & off & 5 & 26.3 & 0.207 & 0.015 & 0.793 & 0.790 & 0.4 \\
blind & progr. & 1 & 29.0 & 0.179 & 0.038 & 0.798 & 0.814 & $-$1.6 \\
blind & progr. & 2 & 31.9 & 0.171 & 0.021 & 0.802 & 0.825 & $-$2.3 \\
blind & progr. & 3 & 31.7 & 0.169 & 0.044 & 0.806 & 0.823 & $-$1.7 \\
blind & progr. & 4 & 24.8 & 0.238 & 0.006 & 0.780 & 0.760 & 2.0 \\
blind & progr. & 5 & 30.6 & 0.176 & 0.028 & 0.809 & 0.819 & $-$0.9 \\
blind & ultra & 1 & 27.2 & 0.221 & 0.010 & 0.799 & 0.777 & 2.2 \\
blind & ultra & 2 & 25.9 & 0.210 & 0.013 & 0.794 & 0.788 & 0.7 \\
blind & ultra & 3 & 29.9 & 0.178 & 0.029 & 0.811 & 0.816 & $-$0.5 \\
blind & ultra & 4 & 29.2 & 0.195 & 0.013 & 0.800 & 0.803 & $-$0.2 \\
blind & ultra & 5 & 27.8 & 0.192 & 0.006 & 0.803 & 0.807 & $-$0.4 \\
reasoning & off & 1 & 37.0 & 0.137 & 0.020 & 0.858 & 0.860 & $-$0.2 \\
reasoning & off & 2 & 32.8 & 0.157 & 0.015 & 0.840 & 0.840 & 0.0 \\
reasoning & off & 3 & 32.8 & 0.165 & 0.024 & 0.828 & 0.831 & $-$0.3 \\
reasoning & off & 4 & 30.8 & 0.169 & 0.013 & 0.829 & 0.829 & 0.1 \\
reasoning & off & 5 & 31.0 & 0.177 & 0.020 & 0.827 & 0.819 & 0.8 \\
reasoning & progr. & 1 & 33.9 & 0.162 & 0.011 & 0.845 & 0.836 & 0.9 \\
reasoning & progr. & 2 & 33.1 & 0.161 & 0.014 & 0.842 & 0.837 & 0.5 \\
reasoning & progr. & 3 & 31.8 & 0.178 & 0.008 & 0.834 & 0.820 & 1.3 \\
reasoning & progr. & 4 & 32.3 & 0.165 & 0.012 & 0.837 & 0.833 & 0.4 \\
reasoning & progr. & 5 & 32.7 & 0.153 & 0.017 & 0.840 & 0.845 & $-$0.5 \\
reasoning & ultra & 1 & 33.8 & 0.156 & 0.022 & 0.841 & 0.841 & 0.0 \\
reasoning & ultra & 2 & 34.9 & 0.148 & 0.019 & 0.846 & 0.849 & $-$0.3 \\
reasoning & ultra & 3 & 33.4 & 0.155 & 0.017 & 0.841 & 0.842 & $-$0.1 \\
reasoning & ultra & 4 & 31.9 & 0.171 & 0.009 & 0.835 & 0.827 & 0.8 \\
reasoning & ultra & 5 & 34.9 & 0.149 & 0.023 & 0.847 & 0.847 & 0.0 \\
\midrule
\multicolumn{3}{@{}l}{all 30 sessions} & & 0.14--0.24 & 0.004--0.044 & 0.78--0.86 & 0.76--0.86 & 0.0 ($-$2.3 to 2.2) \\
blind & ultra & 30 Sept. & 34.8 & 0.154 & 0.012 & 0.819 & 0.844 & $-$2.5 \\
\bottomrule
\end{tabular}
}{\textcolor{red}{fig/validation\_table.tex missing}}}
\end{table}

\begin{figure}[p]
\centering
\includegraphics[width=\linewidth]{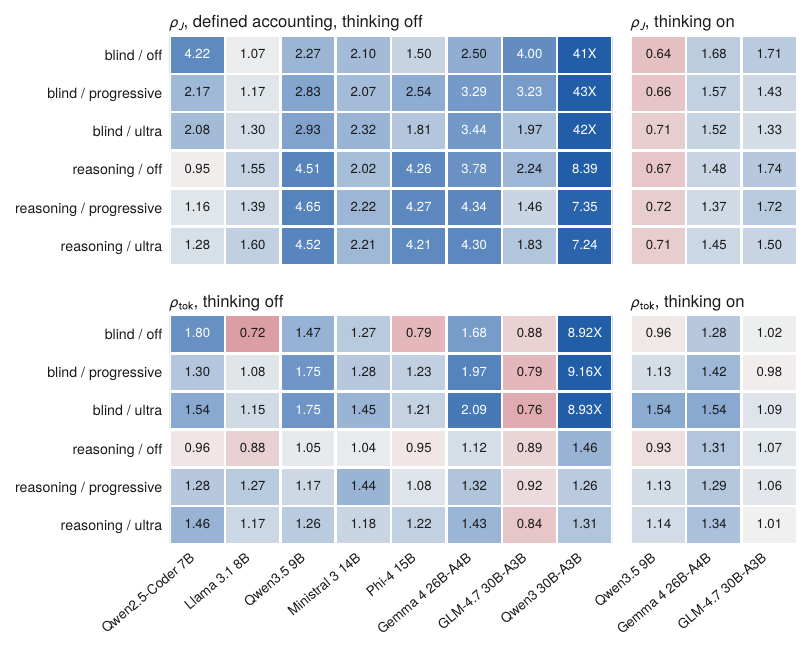}
\caption{The joule ratio under the defined accounting (top) and the token ratio
(bottom) per cell, geometric means: the main set with thinking off (left, eight pairs)
and the three pairs of the treatment arm with thinking on (right). Blue cells are
cells where the attacker spends more than the defender ($\rho > 1$), red cells where
the defender spends more; the colour scale is logarithmic and centred on one. X
marks the truncated blind attacker of Qwen3 30B-A3B.}
\label{fig:s1}
\end{figure}

\begin{figure}[p]
\centering
\includegraphics[width=\linewidth]{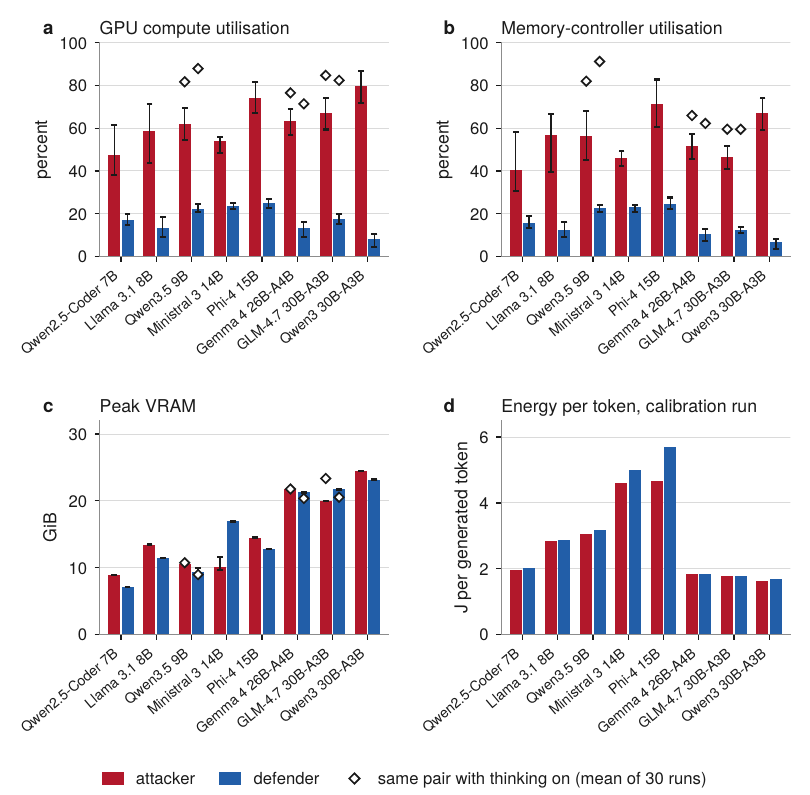}
\caption{Resource profile of the two sides per pair on the main set (thinking off):
(a) GPU compute utilisation and (b) memory-controller utilisation, means over the
session as reported by nvidia-smi; (c) peak VRAM; (d) the energy per generated token
of the calibration run of continuous generation before the campaign, from the
measured load and idle power and the generation rate. Bars are the mean over the six
cells of a pair, whiskers the range of the cell means; diamonds give the same pair with
thinking on. The per-token cost of the two machines agrees within 5 percent in six
pairs of eight, and where it differs, by 8 percent for Ministral and 22 percent for
Phi-4, it is the defender's machine that costs more per token; the joule asymmetry of
the sessions, which runs the other way, therefore comes from how the two sides use the
GPU, utilisation and the balance of prompt reading against generation, not from a
difference in hardware.}
\label{fig:s2}
\end{figure}

\end{document}